\documentclass[useAMS,usenatbib,articles] {mn2e}
\usepackage{epsfig}
\usepackage{graphicx}
\usepackage{amssymb} 
\newcommand{\ratio}{\mbox{$v_{\infty}$/$v_{\rm esc}$}}
\newcommand{\teff}{\mbox{$T_{\rm eff}$}}
\newcommand{\zsun}{\mbox{$Z_{\odot}$}}

\newcommand{\vinf}{\mbox{$v_{\infty}$}}

\title[star formation efficiency-metallicity relation]{Star formation efficiency as a function of metallicity: \\ from star clusters to galaxies}

\author[S. Dib et al.]{Sami Dib$^{1}$\thanks{E-mail: s.dib@imperial.ac.uk}, Laurent Piau$^{2}$,  Subhanjoy Mohanty$^{1}$, Jonathan Braine$^{3}$\\ 
$^{1}$Astrophysics Group, Blackett Laboratory, Imperial College London, London, SW7 2AZ, United Kingdom\\
$^{2}$LATMOS, 11 Boulevard d'Alembert, 78280, Guyancourt, France\\ 
$^{3}$Laboratoire d'Astrophysique de Bordeaux, Universit\'{e} de Bordeaux, OASU CNRS/INSU, 33271, Floirac, France\\}

\begin{document}
\maketitle

\date{Accepted XXX. Received XXX}

\pagerange{\pageref{firstpage}--\pageref{lastpage}}
\pubyear{2009}
\label{firstpage}

\begin{abstract} 

We explore how the star formation efficiency in a protocluster clump is regulated by metallicity dependent stellar winds from the newly formed massive OB stars ($M_{\star} \ge 5~$ M$_{\odot}$) on their main sequence. The model describes the co-evolution of the mass function of gravitationally bound cores and of the IMF in a protocluster clump. Dense cores are generated uniformly in time at different locations in the clump, and contract over lifetimes that are a few times their free fall times. The cores collapse to form stars that power strong stellar winds whose cumulative kinetic energy evacuates the gas from the clump and quenches further core and star formation. This sets the final star formation efficiency, $SFE_{f}$. Models are run with various metallicities in the range $Z/Z_{\odot}=[0.1,2]$. We find that the $SFE_{f}$ decreases strongly with increasing metallicity. The $SFE_{f}$-metallicity relation is well described by a decaying exponential whose exact parameters depend weakly on the value of the core formation efficiency. We find that there is almost no dependence of the $SFE_{f}$-metallicity relation on the clump mass. This is due to the fact that an increase (decrease) in the clump mass leads to an increase (decrease) in the feedback from OB stars which is opposed by an increase (decrease) in the gravitational potential of the clump.The clump mass-cluster mass relations we find for all of the different metallicity cases imply a negligible difference between the exponent of the mass function of the protocluster clumps and that of the young clusters mass function. By normalizing the $SFE$s to their value for the solar metallicity case, we compare our results to $SFE-$metallicity relations derived on galactic scales and find a good agreement. As a by-product of this study, we also provide ready-to-use prescriptions for the power of stellar winds of main sequence OB stars in the mass range [$5,80$] M$_{\odot}$ in the metallicity range we have considered. 

\end{abstract} 


\section{INTRODUCTION}\label{intro}

\subsection{OBSERVATIONS OF THE SFE IN STELLAR CLUSTERS AND IN GALAXIES}

\begin{figure}
\begin{center}
\epsfig{figure=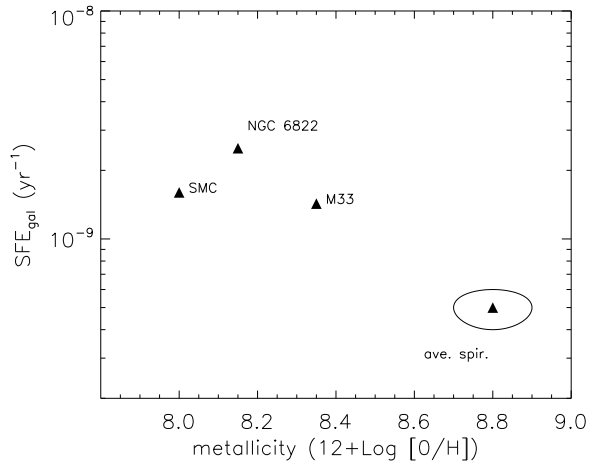,width=\columnwidth}
\end{center}
\caption{Global galactic star formation efficiency $SFE_{gal}$ as a function of the galactic global metallicity. The data for M33 and NGC 6822 are from Gratier et al. (2010a,b), for the SMC from Leroy et al. (2006), and the average spiral value from the sample of Murgia et al. (2002).}
\label{fig1}
\end{figure}

One of the most essential quantities that regulates the dynamical evolution and chemical enrichment of stellar clusters, the interstellar medium, and galaxies is the star formation efficiency (SFE) (e.g., Geyer \& Burkert 2000; Boissier et al. 2001; Krumholz \& McKee 2005; Dib et al. 2006; Parmentier \& Fritze 2009; Dib et al. 2009). The SFE is commonly defined as the fraction of gas which is converted into stars in a system of a given mass, be it a protocluster molecular clump, an entire giant molecular cloud (GMC), or a galaxy. In a non isolated system, such as a protocluster clump embedded in a filamentary structure from which the clump continues to accrete, the final star formation efficiency, $SFE_{f}$, is difficult to define. It can be approximated by:

\begin{equation} 
SFE_{f}=SFE(t=t_{exp})= \frac {M_{cluster}(t_{exp})} {M_{gas,i}+M_{gas,acc}(t_{exp})},
\label{eq1}
\end{equation}

\noindent where $t_{exp}$ is the epoch at which the gas is expelled from the protocluster region, or otherwise depleted. $M_{cluster} (t_{exp})$ is the final stellar mass, $M_{gas,i}$ is the initially available reservoir of star forming gas, and $M_{gas,acc}(t_{exp})=\int_{0}^{t_{exp}} \dot{M}_{cl,acc}(t^{'})~dt^{'}$ is the amount of gas the clump has accreted from the filament and from which also stars form, with $\dot{M}_{cl,acc}$ being the clump's time dependent accretion rate. For a clump with low levels of accretion or for which star formation started after the bulk of the gas in the filament have been accreted onto it, Eq.~\ref{eq1} is reduced to:

\begin{equation} 
SFE_{f} \approx \frac {M_{cluster}} {M_{clump}},
\label{eq2}
\end{equation}

\noindent where $M_{clump}$ is the mass of the protocluser clump, and $M_{cluster}$ is the mass of the stellar cluster. However, in the observations, it is difficult to estimate what the original clump mass was, once a large fraction of the gas has been expelled from the protocluster region. Nevertheless, the SFE can be measured in embedded or semi-embedded clusters and is usually defined as being: 

\begin{equation} 
SFE_{obs}= \frac{M_{cluster}}{M_{cluster}+M_{gas}},
\label{eq3}
\end{equation}

\noindent where $M_{gas}$ is the mass of the star forming gas in the protocluster region. Measured SFEs of nearby embedded clusters, using Eq.~\ref{eq3} yield values that fall in the range 0.1-0.4 (Cohen \& Kuhi 1979; Wilking \& Lada 1983; Rengarajan 1984; Wolf et al. 1990; Pandey et al. 1990; Lada et al. 1991a,1991b; Warin et al. 1996; Olmi \& Testi 2002). These SFE values are larger than those obtained for entire GMCs which are observed to fall in the range of a few percent (e.g., Duerr et al. 1982; Myers et al. 1986, Fukui \& Mizuno 1991; Evans et al. 2009). On galactic scales, the star formation efficiency is usually defined as being the inverse of the molecular gas consumption time, and is given by:

\begin{equation} 
SFE_{gal}= \frac{SFR_{gal}}   {M_{H_{2}}},
\label{eq4}
\end{equation}

\noindent where $SFR_{gal}$ is the galactic star formation rate (in M$_{\odot}$ yr$^{-1}$) and $M_{H_{2}}$ the mass of the molecular hydrogen gas. In some studies, $M_{H_{2}}$ is replaced by $M_{gas}$ which is the total mass of the gas in the galaxy, or by the mass of the H I gas (e.g., Schiminovitch et al. 2010). In large spiral galaxies $SFE_{gal} \sim 0.5\times 10^{-9}$ yr$^{-1}$ (e.g., Kennicutt 1989,1998, Murgia et al. 2002; Leroy et al. 2008). The local SFE in galaxies is found to depend on galactic radius (e.g., Zasov \& Abramova 2006; Leroy et al. 2008). Among spiral galaxies, the global $SFE_{gal}$ is found not to depend clearly on the Hubble type and to depend weakly on the galactic environment (e.g., Young et al. 1986;  Rownd \& Young 1999). However, Young et al. (1996) report that the $SFE_{gal}$ of Irregular galaxies (which are usually less metal-rich) is higher that that of normal spirals (Fig. 11 in their paper). Boissier et al. (2001) found no dependence of the $SFE_{gal}$ on some global galactic properties such as  the circular velocity, the H band luminosity of the galaxies, and their $B-H$ colour index. In Fig.~\ref{fig1} we have compiled the global $SFE_{gal}$ for a few selected nearby galaxies as a function of their global metallicity. The trend in Fig.~\ref{fig1} is strongly suggestive of an $SFE_{g}$ which increases with decreasing metallicity. Interestingly, a recent study by Mannucci et al. (2010) using SDSS data for nearby galaxies in combination with a sample of higher redshift galaxies confirmed the existence, for a given galactic stellar mass, of a dependence of the $SFR_{gal}$ and of the specific star formation rate ($SFR_{gal}$ per unit stellar mass, $SSFR_{gal}$) on metallicity. Their results show that the $SFR_{gal}$ and the $SSFR_{gal}$ increase with decreasing galactic metallicities. A similar result was also pointed out by Ellison et al. (2008) and is also visible when plotting the SSFR as a function of metallicity using the data of Lara-L\'{o}pez et al. (2010).   
 
\subsection{REGULATION OF THE SFE IN STAR FORMING REGIONS}

Several physical processes regulate the core formation efficiency (CFE) within a star forming molecular clump/cloud, and as a consequence, the SFE. Supersonic turbulence that is ubiquitously observed in molecular clouds can, if driven, provide support against the global collapse of the clump/cloud, or at least delay it (e.g., V\'{a}zquez-Semadeni \& Passot 1999). However, on scales smaller than the energy injection scale, but larger than the sonic scales in the cloud, supersonic turbulence produces local compressions (cores) (e.g., Padoan 1995; V\'{a}zquez-Semadeni et al. 2003) of which a fraction can be 'captured' by gravity and proceed to collapse into stars (e.g., V\'{a}zquez-Semadeni et al. 2005a; Dib et al. 2007a; Dib \& Kim 2007; Dib et al. 2008a; Dib et al. 2010a). Work by Clark \& Bonnell (2004) and Klessen et al. (2000) suggested that the CFE decreases with increasing turbulent energy and decreasing energy injection scale. Krumholz \& McKee (2005) formulated an analytical theory in which the star formation rate per free-fall time, $SFR_{ff}$, is shown to decrease with an increasing sonic Mach number and an increasing virial parameter (see also the simulation of Clark \& Bonnell 2004). Padoan \& Nordlund (2011) argued that the $SFR_{ff}$ increases with increasing Mach number and that it also depends weakly on the ratio of the mean gas pressure of the cloud to the mean magnetic pressure. A similar trend for the dependence of the $SFR_{ff}$ on the sonic Mach number was also reported by Rosas-Guevara et al. (2010), who additionally showed that the mass contained in colliding streams from which protocluster clumps form is also a key parameter in regulating the fraction of the gas that can become gravitationally bound.  

Magnetic fields also play an important role in determining the fraction of gravitationally bound gas in star forming clouds/clumps. Results from both ideal and non-ideal MHD simulations show that stronger magnetic fields (in terms of magnetic criticality) lower the rate of dense core formation in a star forming molecular clump/cloud (e.g., V\'{a}zquez-Semadeni et al. 2005b; Price \& Bate 2008; Dib et al. 2008a; Li et al. 2010; Dib et al. 2010a). Dib et al. (2010a) showed that the CFE per unit of the free-fall time of the cloud, $CFE_{ff}$, are of the order of $\sim 6~\%$ and $\sim 33~\%$ for clouds with mass to-magnetic flux ratios of $\mu_{B}=2.2$ and $8.8$, respectively (with $\mu_{B}$ being normalized by the critical mass-to-flux ratio for collapse).  

Albeit magnetic fields and turbulence regulate the rate at which gas in a cloud/clump is turned into dense cores, and subsequently into stars, the SFE in a star forming region will continue to increase over time until either the star forming gas is exhausted or otherwise expelled from the cloud/clump. The role of stellar feedback in setting the final value of the SFE by totally removing the gas has been investigated by several authors. The role of protostellar outflows has been studied theoretically by Adams \& Fatuzzo (1996), Matzner \& McKee (2000), Matzner (2002,2007) and numerically by Nakamura \& Li (2005,2007), Li \& Nakamura (2006) and Li et al. (2010). Although protostellar outflows seem to play an important role in generating a self-sustained turbulence in a protocluster forming region, they do not inject enough energy that can lead to the gas removal from the region (e.g., Nakamura \& Li 2007). Supernova explosions are an efficient way of removing gas from the protocluster region (e.g., Parmentier et al. 2008; Baumgardt et al. 2008). However, such explosions occur after a few million years from the time massive stars have formed (i.e., the hydrogen burning phase of a 9,  a 20, and a 40 M$_{\odot}$ solar metallicity stars are $\gtrsim 22, 7$, and $4$ Myr, respectively; Meynet \& Maeder 2000). Thus, supernova may play a role in the removal of the gas only in regions where other gas removal mechanisms would be inefficient due for example to a very low core and star formation efficiencies. 

Another form of stellar feedback is associated with O and B stars, in their main sequence phase, and eventually beyond. OB stars emit UV radiation which ionizes the surrounding gas and heats it to temperatures of $\sim 7000-10^{4}$ K. This warm and ionized bubble provides the environment in which particles accelerated from the stellar surface by interaction with some of the stellar radiation propagate outwards. In the first few hundred years, the wind freely expands at the wind velocity until it shocks with the surrounding material and/or with the wind of a neighbouring massive star. Shocked material is brought to temperatures of $10^{6}-10^{7}$ K. The hot gas expands supersonically in the warm medium. When the densities of the swept-up material reach a high value, a cooling instability operates behind the shock front leading to the formation of a thin dense shell. Models of energy driven wind bubbles which take into account the interaction of the expanding bubble with a surrounding confining medium were developed by Castor et al. (1975), Weaver et al. (1977), Shull (1980), Koo \& McKee (1992a,b) and Cant\'{o} et al. (2000), and recently included in global molecular cloud simulations by V\'{a}zquez-Semadeni et al. (2010). In contrast, models developed by Chevalier \& Clegg (1985) and Stevens \& Hartwell (2003) ignored the surrounding material and described the evolution of a free-floating wind. McKee et al. (1984) pointed out that the hot gas pressure in the bubble can be reduced by leakage of energy in the form of radiation. Harper-Clark \& Murray (2009) argued that since the winds are likely to be clumpy (e.g., L\'{e}pine \& Moffat 2008; Prinja \& Massa 2010), the pressure in the bubble is reduced by the escape of the hot gas rather than the escape of radiation. 

For star clusters whose masses are $\lesssim 10^{4}$ M$_{\odot}$ (corresponding to luminosities of $\sim 10^{50}$ ionizing photon s$^{-1}$ for a fully sampled IMF with a Salpeter slope), the direct contribution of radiation pressure in the evacuation of gas from around the star/star cluster is minimal (e.g., Mathews 1969; Gail \& Sedlmayr 1979; Arthur et al. 2004; Henney 2007; Krumholz \& Matzner 2009).  For more massive clusters, like those found in the Antennae, M82, and potentially some of the Milky Way starburst clusters such as Arches, several authors have recently pointed out that direct radiation pressure might be the dominant mode of gas expulsion from the protocluster region (Krumholz \& Matzner 2009; Fall et al. 2010; Murray et al. 2010). 
 
\subsection{AIM OF THIS WORK}\label{aim}

An entirely unexplored area in star formation theories is the dependence of the star formation efficiency in a protocluster environment on metallicity. As stellar winds are strongly metallicity dependent (e.g., Vink et al. 2001; Bresolin \& Kudritzki 2004), feedback by winds is expected to lead to variations in the SFE in protocluster regions which are identical except in the metallicity of the gas. To investigate this effect, we use a modified version of the model developed by Dib et al. (2010b) which describes the co-evolution of the mass function of gravitationally bound cores and the IMF in a protocluster region. In this model, gas is evacuated from the protocluster clump by the stellar winds of the newly formed massive stars. In order to focus on the effects of the metallicity dependent feedback from stellar winds, we neglect in this paper the evolution of the core mass function (CMF) under the effects of clump gas accretion by the cores and core coalescence. The effects of these processes have been explored by  Shadmehri 2004, Basu \& Jones 2004; Dib 2007, Dib et al. 2007b,2008b,2010b; Veltchev et al. 2010. In the Dib et al. (2010b) model, dense cores form in the protocluster clump uniformly in time following a specified core formation efficiency per unit free-fall time of the clump, $CFE_{ff}$. The dense cores have lifetimes of a few times their free-fall times, after which they collapse to form stars and are removed from the CMF. Stellar winds from the newly formed massive stars ($M_{\star} > 5~$M$_{\odot}$) inject energy into the protocluster clump. In order to calculate the stellar wind luminosities, we use a state of the art modified version of the stellar evolution code CESAM (Morel 1997, Morel \& Lebreton 2008, Piau et al. 2011) to calculate main sequence models for stars in the mass range [5-80] M$_{\odot}$ and derive their basic stellar properties; i.e., luminosity, effective temperature, and radius. The stellar grids are calculated for a range of metallicities between $Z=1/10~Z_{\odot}$ and $Z=2~Z_{\odot}$. In a second step, we use the stellar atmosphere model of Vink et al. (2001) in order to calculate the mass loss rates and terminal velocities of the winds. 

In \S.~\ref{model}, we describe the protocluster clump model, the dense core model, and the distributions of cores that form in the clump (i.e., the initial CMF). In \S.~\ref{feedback}, we describe the metallicity dependent stellar wind models and the derivation of the stellar wind luminosities. The co-evolution of the prestellar core mass function and of the IMF is presented in detail for a fiducial case with solar metallicity in \S.~\ref{co_evol_fiducial}. We also show in this section how the feedback from stellar winds helps define the final SFE in the protocluster region, $SFE_{f}$. The dependence of the of SFE on metallicity is presented in \S.~\ref{metal_dep} for the fiducial clump mass. In \S.~\ref{mass_dep} we investigate the dependence of $SFE_{f}$-metallicity relation on the clump mass. In \S.~\ref{m_m_relation}, we discuss the implication of our results on the difference between the exponents of the clump mass function and the young clusters mass function and in \S.~\ref{impli_galaxies} we discuss the implications of our results for the SFE-metallicty relation on galactic scales as well as their implications for the SFE in globular clusters. In \S.~\ref{summary}, we summarize our results.       

\begin{figure}
\begin{center}
\epsfig{figure=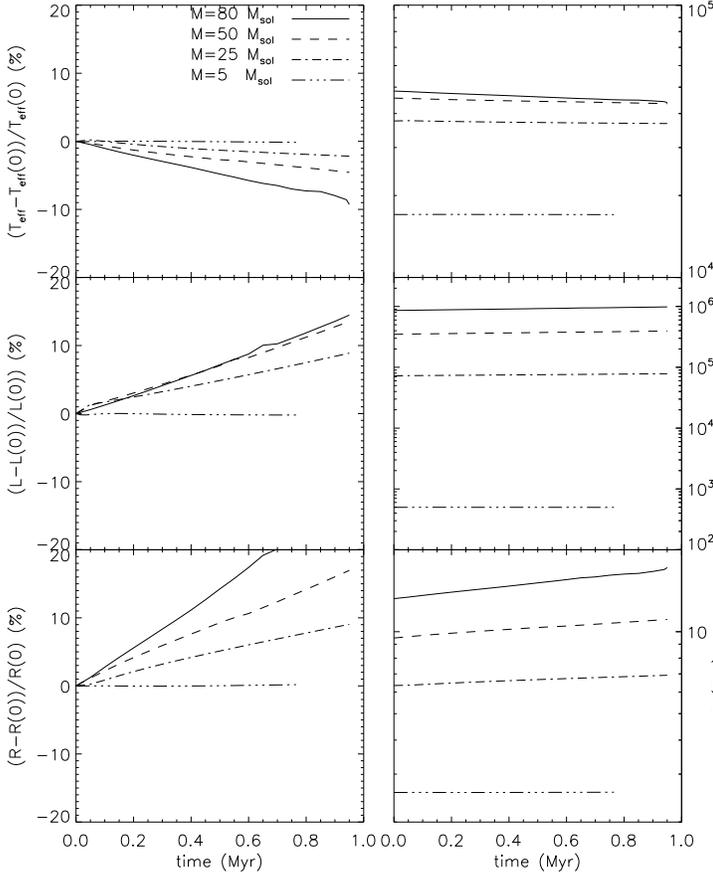,width=\columnwidth}
\end{center}
\vspace{0.5cm}
\caption{Time evolution of the effective temperature (top), luminosity (middle), and radius (bottom) for solar metallicity stars of a few selected masses evolving on the main sequence. The calculation have been performed using the stellar evolution code CESAM (Morel 1997; Morel \& Lebreton 2008). The three quantities are shown in real values (right column) and in percentage of variation with respect to the initial value (left column). }
\label{fig2}
\end{figure}

\section{THE CLUMP AND CORE MODELS}\label{model}

\subsection{PROTOCLUSTER CLUMPS: OBSERVATIONS AND MODELS}\label{clusters}

Several studies have established that star clusters form in dense ($\gtrsim 10^{3}$ cm$^{-3}$) clumps embedded in a lower density parental molecular cloud (e.g., Lada \& Lada 2003; Lada et al. 2010; Csengeri et al. 2011; Parmentier 2011). Saito et al. (2007) recently studied, using  the C$^{18}$O molecular emission line, a large sample of cluster forming clumps whose masses and radii vary between [15-1500] M$_{\odot}$Êand [0.14-0.61] pc, respectively. Based on the sample of clumps of Saito et al. (2007), the mass-size, and velocity dispersion-size relations were fitted by Dib et al. (2010b) and are given by: 

\begin{equation}
M_{clump}({\rm M_{\odot}})=10^{3.62 \pm 0.14} R_{c}^{2.54 \pm 0.25} ({\rm pc}), 
\label{eq5}
\end{equation}

\noindent and 

\begin{equation}
v_{c}({\rm km~s^{-1}})=10^{0.45 \pm 0.08} R_{c}^{0.44 \pm 0.14}({\rm pc}).
\label{eq6}
\end{equation}

In this work, we adopt a protocluster clump model that follows an $r^{-2}$ density profile:

\begin{equation} 
\rho_{c}(r)= \frac{\rho_{c0}} {1+(r/R_{c0})^{2}},
\label{eq7}
\end{equation}

\noindent where $R_{c0}$  is the clump's core radius (core radius here stands for the central region of the clump), $\rho_{c0}$ is the density at the centre. In the study of Mueller et al. (2002), the average slope of the density profiles of 51 star forming clumps is $-1.8 \pm 0.4$ (here we adopt a value of $-2$). For a given mass $M_{clump}$ of the clump, the central density $\rho_{c0}$ is given by:

\begin{equation} 
\rho_{c0}= \frac{M_{clump}}{4 \pi R_{c0}^{3} [(R_{c}/R_{c0})-\arctan(R_{c}/R_{c0})]},
\label{eq8}
\end{equation}

\noindent where $R_{c}$ is the size of the clump. The temperatures of the cluster forming clumps are observed to vary between 15 and 70 K (e.g., Saito et al. 2007). In order to further constrain the models and minimize the number of parameters, we relate the sizes of the protocluster clumps to their masses using the mass-size relation of Saito et al.  In the absence of detailed information about the velocity dispersion inside the clumps in the Saito et al. (2007) study, we assume that the clump-clump velocity dispersion they derived (i.e., Eq~\ref{eq6}) is also valid on the scale of the clumps themselves and of their substructure. 

\subsection{THE PRESTELLAR CORE MODEL}\label{core_model}

Whitworth \& Ward-Thompson (2001) applied a family of Plummer sphere-like models to the contracting prestellar dense core L1554, which is representative of the population of gravitationally bound cores in clumps that are considered in this work. They found a good agreement with the observations of L1554 if the density profile of the core has the following form:

\begin{equation} 
\rho_{p}(r_{p})= \frac{\rho_{p0}}{[1+(r_{p}/R_{p0})^{2}]^{2}},
\label{eq9}
\end{equation}

\noindent where $\rho_{p0}$ and $R_{p0}$ are the central density and core radius of the core, respectively. Note that the radius of the core, $R_{p}$, depends both on its mass and on its position within the clump. The dependence of $R_{p}$ on $r$ requires that the density at the edges of the core equals the ambient clump density, i.e., $\rho_{p}(R_{p})=\rho_{c}(r)$. This would result in smaller radii for cores of a given mass when they are located in their inner parts of the clump. The density contrast between the centre of the core and its edge is given by: 

\begin{equation} 
{\cal C}(r) \equiv \frac {\rho_{p0}}{\rho_{c} (r)}=\frac {\rho_{p0}} {\rho_{c0}} \left[1+ \left(\frac{r}{R_{c0}}\right)^{2} \right].         
\label{eq10}
\end{equation}

Depending on its position $r$ in the clump, the radius of the core of mass $M$, $R_{p}$, can be calculated as being $R_{p} (r,M)=a(r)~R_{p0} (r,M)$, where: 

\begin{equation} 
R_{p0}(r,M)= \left(\frac{M}{2 \pi \rho_{p0}} \right)^{1/3} \left(\arctan[a(r)]-\frac{a(r)}{1+a(r)^{2}} \right)^{-1/3},
\label{eq11}
\end{equation} 

\noindent and with $a(r) \equiv ({\cal C}(r)^{1/2}-1)^{1/2}$. With our set of parameters, the quantity ${\cal C}^{1/2}-1$ is always guaranteed to be positive. The value $R_p(r,M)$ can be considered as being the radius of the core at the moment of its formation. The radius of the core will decrease as time advances due to gravitational contraction. We assume that the cores contract on a timescale, $t_{cont,p}$ which we take to be a few times their free fall timescale $t_{ff}$, and which is parametrized by $t_{cont,p}(r,M)= \nu ~ t_{ff}(r,M)= \nu \left( 3 \pi/32~G \bar{\rho_{p}} (r,M) \right)^{1/2}$, where $G$ is the gravitational constant, $\nu$ is a constant $\ge 1$ and $\bar{\rho_{p}}$ is the radially averaged density of the core of mass $M$, located at position $r$ in the clump. The time evolution of the radius of a core of mass $M$, located at position $r$ in the cloud is given by a simple contraction law $R_{p}(r,M,t)=R_p(r,M,0)~e^{-(t/t_{cont,p})}$. Theoretical considerations suggest that $t_{cont,p}$ can vary between $t_{ff}$ ($\nu=1$) and $10~t_{ff}$ ($\nu=10$) with the latter value being the characteritic timescale of ambipolar diffusion (McKee 1989; Fiedler \& Mouschovias 1992; Ciolek \& Basu 2001). Both observational (Lee \& Myers 1999; Jessop \& Ward-Thompson 2000; Kirk et al. 2005; Hatchell et al. 2007; Ward-Thompson et al. 2007) and numerical (V\'{a}zquez-Semadeni et al. 2005a; Galv\'{a}n-Madrid et al. 2007; Dib et al. 2008c) estimates of gravitationally bound cores lifetimes tend to show that they are of the order of a few times their free-fall time, albeit decreasing (but still larger than one free-fall time) when cores are defined with increasingly higher density tracers/thresholds

One important issue is the choice of the cores central density, $\rho_{p0}$. In this work, we first assume that the minimum density contrast that exists between the centre of the core and its edge is of the order of the critical Bonnor-Ebert value and that is $\gtrsim 15$. Secondly, we assume that the density contrast between the centre and the edge of the cores depends on their masses following a relation of the type $\rho_{p0}  \propto M^{\mu}$. Thus, the density contrast between the centre and the edge for a core with the minimum mass we are considering, $M_{min}$ (typically $M_{min}=0.1$ M$_{\odot}$) is 15, whereas for a more massive core of mass $M$, the density contrast will be equal to $15\times (M/M_{min})^{\mu}$.  Observations show that  $\mu$ varies in the range $[0-0.6]$ (e.g.,  Caselli \& Myers 1995; Johnstone \& Bally 2006).
 
\subsection{THE INITIAL CORE MASS FUNCTION} \label{ini_cond}

We assume that dense cores form in the clump as a result of its gravo-turbulent fragmentation. We describe the mass distribution of cores formed at each epoch using the formulation of Padoan \& Nordlund (2002). Thus, the local mass distributions of cores, $N(r,M)$, are given by: 

\begin{eqnarray}
N (r,M)~d\log~M =f_{0}(r)~M^{-3/(4-\beta)} \nonumber \\
            \times \left[\int^{M}_{0} P(M_{J}) dM_{J}\right]d\log~M,
\label{eq12}
\end{eqnarray}   

\noindent where $\beta$ is the exponent of the kinetic energy power spectrum, $E_{k} \propto k^{-\beta}$, and is related to the exponent $\alpha$ of the size-velocity dispersion relation in the clump with $\beta=2 \alpha+1$.  The local normalization coefficient $f_{0}(r)$ is obtained by requiring that $\int^{M_{max}}_{M_{min}} N (r,M)~dM=1$ in a shell of width $dr$, located at distance $r$ from the clump's centre. $P(M_{J})$ is the local distribution of Jeans masses given by:

\begin{equation}
P(M_{J})~dM_{J}=\frac{2~M_{J0}^{2}}{\sqrt{2 \pi \sigma^{2}_{d}}} M^{-3}_{J} \exp \left[-\frac{1}{2} \left(\frac{\ln~M_{J}-A}{\sigma_{d}} \right)^{2} \right] dM_{J},
\label{eq13}
\end{equation}   

\noindent where $M_{J0}$ is the Jeans mass at the mean local density, and $\sigma_{d}$ is the standard deviation of the density distribution which is a function of the local thermal rms Mach number. Therefore, the local distribution of cores generated in the clump, at an epoch $\tau$, $N(r,M,\tau)$, is obtained by multiplying the local normalized function $N(r,M)$ by the local rate of fragmentation such that:

\begin{equation}
{N} (r,M,\tau) dt=\frac{CFE_{ff} (r) \rho_{c}(r)} {<M>(r)~t_{cont,p} (r,M)} \frac{dt}{t_{ff,cl}} N(r,M),
\label{eq14}
\end{equation} 

\noindent where $dt$ is the time interval between two consecutive epochs, $<M>$ is the average core mass in the local distribution and is calculated by $<M>=\int_{M_{min}}^{M_{max}} M~N (r,M,0)~dM$, and $CFE_{ff}$ is a parameter smaller than unity which describes the local mass fraction of gas that is transformed into cores per free fall time of the protocluster clump, $t_{ff,cl}$. In the present study, we assume that $CFE_{ff}$ is independent of $r$. 
  
\section{FEEDBACK MODEL: METALLICITY DEPENDENT STELLAR WINDS OF OB STARS}\label{feedback}

\begin{figure}
\begin{center}
\epsfig{figure=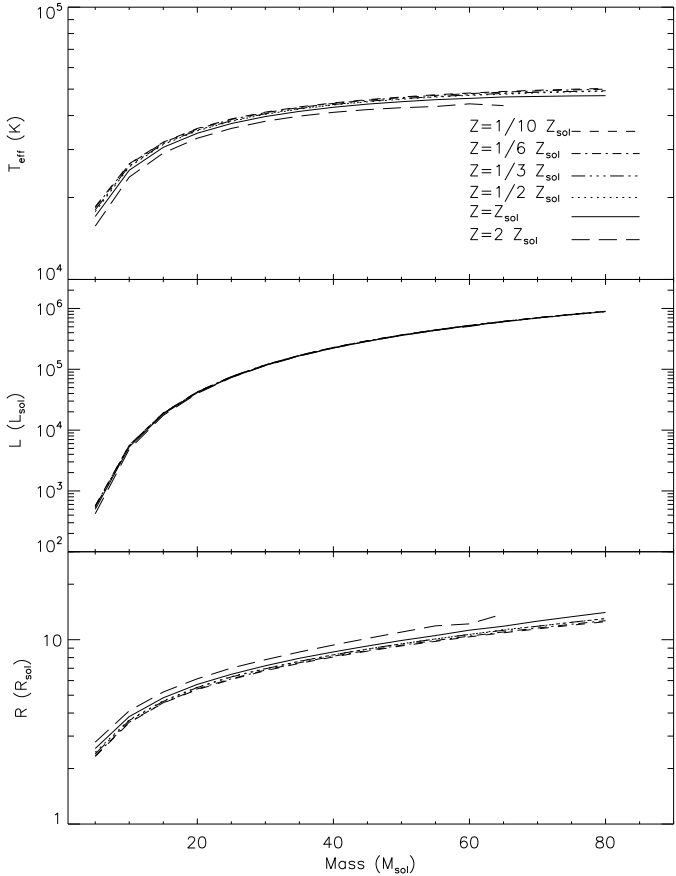,width=\columnwidth}
\end{center}
\vspace{0.5cm}
\caption{Time averaged, over the first $5 \times 10^{5}$ yr of their life on the main sequence, effective temperature (top), luminosity (middle), and radius (bottom) for stars of masses between 5 and 80 M$_{\odot}$ and metallicities ranging from $Z=1/10~Z_{\odot}$ to $Z=2~Z_{\odot}$.}
\label{fig3}
\end{figure}

We assume that the formation of cores in the protocluster clump, and consequently star formation, is terminated whenever the fraction of the wind energy stored into motions that oppose gravity exceeds the gravitational energy of the clump. Whenever this occurs (at epoch $t=t_{exp}$), the gas is expelled from the protocluster clump and star formation is quenched.  Thus, at any epoch $t < t_{exp}$, gas is removed from the clump only to be turned into stars. We take into account the feedback generated by the stellar winds of massive stars ($M_{\star} \ge 5$ M$_{\odot}$). In order to calculate reliable estimates of the feedback generated by metallicity dependent stellar winds, we proceed in two steps. 

In the first step, we use a modified version of the stellar evolution code CESAM (see appendix 1 in Piau et al. 2011) to calculate a grid of main sequence stellar models for stars in the mass range [5-80] M$_\odot$ (with steps of 5 M$_{\odot}$) at various metallicities which are $Z/Z_{\odot}=[1/10, 1/6, 1/3, 1/2, 1, 2]$. Provided the solar photosphere metal repartition recently determined by Asplund et al. (2005) is considered, the heavy elements (i.e., elements heavier than hydrogen and helium) fraction results from our latest solar calibrated models (Piau et al. 2011) and yields $Z_{\odot}=0.0138$. In addition to the solar metallicity case, the sample of explored metallicities is selected such that it covers a broad range of observed metallicities in nearby galaxies. This range covers the characteristic metallicities of nearby galaxies such as the Large Magellanic Cloud and M33 $(Z/Z_{\odot} \sim 1/3-1/2$ (e.g, Russell \& Dopita 1992; Braine et al. 2001; Rolleston et al. 2002; Smith et al. 2002), the metallicity of the Small Magellanic Cloud (SMC) $Z_{SMC}=1/8-1/5~Z_{\odot}$ (Russel \& Dopita 1992; Rolleston et al. 2003; Lee et al. 2005; P\'{e}rez-Montero \& D\'{i}az 2005), and metallicities characteristic of metal rich environments such as those prevailing in starburst galaxies such as M 83 or in the circumnuclear starburst regions in the Galaxy and in other galaxies, $Z=1.5-2~Z_{\odot}$ (Zaritsky et al. 1994; Kobulnicky et al. 1999, Bresolin \& Kennicutt 2002). We always consider the solar metal repartition of Asplund et al. (2005) as we deal with typical population I stars. 

The evolution of massive stars was followed using the CESAM code for 1Myr (or slightly less in some cases), on the main sequence. The characteristic stellar properties, which are the effective temperature $T_{eff}$, the luminosity $L_{\star}$, and the stellar radius $R_{\star}$ were dumped every $5 \times 10^{4}$ yr. For ages $< 1$ Myr, all the stars, including the most massive ones, remain on the main sequence. Fig.~\ref{fig2} displays the time evolution of $T_{eff}$, $L_{\star}$, and $R_{\star}$ in the models for a few selected stellar masses for the solar metallicity case, both in physical units (right column), and in the percentage of variation with respect to the initial value (left column). These figures show that the time variations of these quantities are of the order of a few percent, increasing with time to about 10-20 percent at most for the most massive stars. Since we are interested in the early time evolution of proto-clusters in which winds are injecting energy, we average the stellar properties over the first $5 \times 10^{5}$ yr and use the averaged stellar properties as characteristic values for stars on the main sequence. This procedure has been repeated for all other metallicities. The resulting characteristic stellar properties are shown in Fig.~\ref{fig3} for the various metallicity cases. One can notice that for a given mass over the considered mass range, the radius of a higher metallicity star is larger and its effective temperature smaller thus resulting in luminosity values that are almost independent of metallicity.  

In a second step, we use the grid of calculated time averaged stellar properties to evaluate, for the different metallicity cases, the stellar mass loss rates and the power of the stellar winds. To that purpose, we use the results of the stellar atmosphere models developed by Vink et al. (1999,2000,2001). These models allow us to evaluate the stellar mass loss rate $\dot{M}_{\star}$, as a function of the stellar mass $M_{\star}$, effective temperature $T_{eff}$, the stellar luminosity $L_{\star}$, the metallicity $Z$, and the ratio of the velocity of the wind at infinity to the escape velocity, $\ratio$, using the following formulations (Vink et al. 2001):

\begin{eqnarray}
{\rm log}~\dot{M}_{\star} & = &~-~6.697~(\pm 0.061) \nonumber \\
                  & &~+~2.194~(\pm 0.021)~{\rm log}(L_{\star}/{10^5}) \nonumber \\
                  & &~-~1.313~(\pm 0.046)~{\rm log}(M_{\star}/30) \nonumber\\
                  & &~-~1.226~(\pm 0.037)~{\rm log}\left(\frac{\ratio}{2.0}\right) \nonumber \\
                  & &~+~0.933~(\pm 0.064)~{\rm log}(\teff/40 000) \nonumber\\
                  & &~-~10.92~(\pm 0.90)~\{{\rm log}(\teff/40 000)\}^{2} \nonumber\\
                  & &~+~0.85~(\pm 0.10)~{\rm log}(Z/\zsun) \nonumber\\
                  \nonumber\\
                  & &~{\rm for}~27~500 < \teff \le 50000 {\rm K},
\label{eq15}
\end{eqnarray}

\noindent and 

\begin{eqnarray}
{\rm log}~\dot{M}_{\star} & = &~-~6.688~(\pm 0.080) \nonumber \\
                  & &~+~2.210~(\pm 0.031)~{\rm log}(L_{\star}/{10^5}) \nonumber \\
                  & &~-~1.339~(\pm 0.068)~{\rm log}(M_{\star}/30) \nonumber\\
                  & &~-~1.601~(\pm 0.055)~{\rm log}\left(\frac{\ratio}{2.0}\right) \nonumber \\
                  & &~+~1.07~(\pm 0.10)~{\rm log}(\teff/20 000) \nonumber\\
                  & &~+~0.85~(\pm 0.10)~{\rm log}(Z/\zsun) \nonumber\\
                  \nonumber\\
                  & &~{\rm for}~12500 \le \teff \le 22500 {\rm K}
\label{eq16}                  
\end{eqnarray}

\noindent where $L_*$ and $M_*$ are in solar units and $\teff$ is in Kelvin. In this range the Galactic ratio of $\ratio$ = 2.6 (Vink et al. 2001). If the values for $\vinf$ at other metallicities are different 
from these Galactic values, then the mass-loss rates can easily be scaled accordingly. The escape velocity of the gas at the surface of the star can be easily calculated using the standard formula:

\begin{equation}
v_{esc}=\left( \frac{2~G~M_{\star,eff}}{R_{\star}} \right)^{1/2}, 
\label{eq17}
\end{equation}

\noindent where $G$ is the gravitational constant and $M_{\star,eff}$ is the effective stellar mass $M_{\star,eff}=M_{\star} (1-H)$, where $H$ is the factor that takes into account the correction of newtonian gravity by the radiation pressure due to electron scattering and which is given by (e.g., Lamers \& Leitherer 1993): 

\begin{equation}
H=7.66 \times 10^{-5} \sigma_{e} \left(\frac{L_{\star}} {L_{\odot}}\right)  \left(\frac{M_{\odot}} {M_{\star}}\right),
\label{eq18}
\end{equation}

\noindent where $\sigma_{e}$ is the electron scattering coefficient, which is given, for compositions in which the number of Helium and Hydrogen atoms follows $n(He)/(n(He)+n(H)) \sim 0.1$, by $\sigma_{e}=0.34$ cm$^{2}$ g$^{-1}$ if $T_{eff} \ge 35000$ K,  $\sigma_{e}=0.32$ cm$^{2}$ g$^{-1}$ when 30000 K $\leq T_{eff}  < 35000$ K, and $\sigma_{e}=0.31$ cm$^{2}$ g$^{-1}$  for $T_{eff}  <  30000$ K (Pauldrach et al. 1990). Vink et al. (2001) did not derive the values of $v_{\infty}$, therefore, we use instead the relations obtained by Leitherer et al. (1992) in which $v_{\infty}$ is given by:

\begin{eqnarray}
{\rm log}~v_{\infty} ({\rm km s^{-1}}) & = &~~1.23-0.30~{\rm log} \left(\frac{L_{\star}}{L_{\odot}}\right) \nonumber \\ 
      & &~+~0.55~{\rm log} \left(\frac {M_{\star}}{M_{\odot}}\right) \nonumber \\
      & &~+~0.64~{\rm log} (T_{eff}) \nonumber \\
      & &~+~0.13 ~{\rm log} \left(\frac{Z}{Z_{\odot}}\right)
\label{eq19}      
\end{eqnarray}

\begin{figure}
\begin{center}
\epsfig{figure=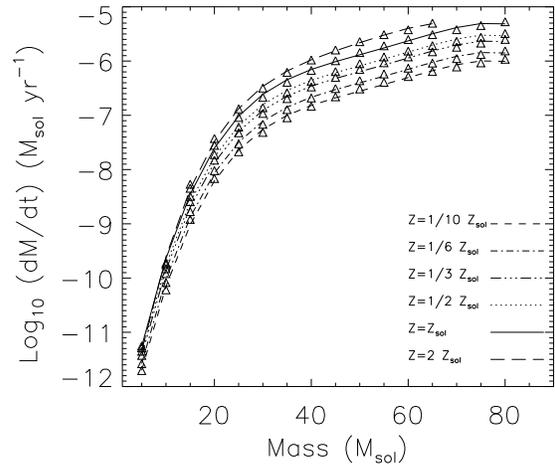,width=\columnwidth}
\end{center}
\caption{Stellar mass loss rates for stars in the mass range 5-80 M$_{\odot}$ on the main sequence, and for various metallicities. The stellar mass loss rates have been calculated using the stellar characteristics (effective temperature, stellar luminosity and radius) computed using the stellar evolution code CESAM coupled to the stellar atmosphere model of Vink et al. (2001). Over-plotted to the data are fourth order polynomial fits. The parameters of the fit functions are summarized in Tab.~\ref{tab1}}
\label{fig4}
\end{figure}

\begin{figure}
\begin{center}
\epsfig{figure=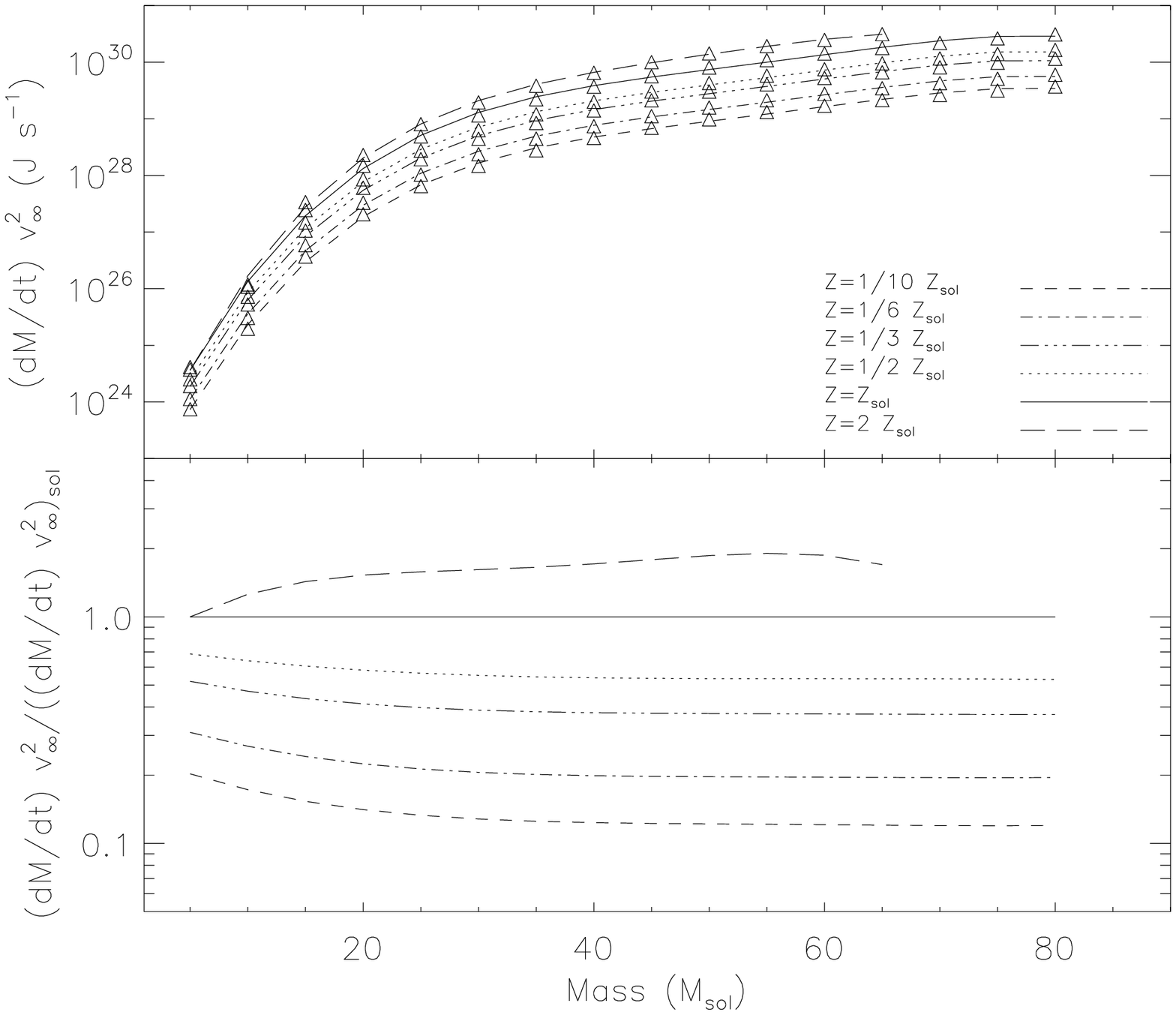,width=\columnwidth}
\end{center}
\vspace{0.7cm}
\caption{The power of the stellar winds, or wind luminosities, for stars in the mass range 5-80 M$_{\odot}$ on the main sequence, and for various metallicities. The stellar mass loss rates have been calculated using the stellar characteristics (effective temperature, stellar luminosity and radius) computed using the stellar evolution code CESAM coupled to the stellar atmosphere model of Vink et al. (2001). The values of $v_{\infty}$ have been calculated using the derivation by Leitherer et al. (1992). Over-plotted to the data are fourth order polynomials. The parameters of the fit functions are summarized in Tab.~\ref{tab2}.}
\label{fig5}
\end{figure}

Fig.~\ref{fig4} displays the mass loss rates calculated for the various metallicities, using equations \ref{eq15}-\ref{eq19} for OB stars in the mass range $[5,80]$ M$_{\odot}$ (triangles). In order to allow the calculation of mass loss rates for a star of any given mass, the data points in Fig.~\ref{fig4} were fitted using fourth order polynomials of the form ${\rm log} (\dot{M}_{\star})=\sum_{i=0,4} A_{i}~M_{\star}^{i}$ for each of the considered metallicities. The values of the coefficients $A_{i}$, are reported in Tab.~\ref{tab1}. The power of the stellar winds is given by $\dot{M}_{\star} v_{\infty}^{2}$. This quantity is displayed in Fig.~\ref{fig5} for the models with different metallicities and the data points are fitted with a fourth order polynomial ${\rm log} (\dot{M}_{\star}~v_{\infty}^{2})=\sum_{i=0,4} B_{i}~M_{\star}^{i}$. The coefficients $B_{i}$ for the models with different metallicities are summarized in Tab.~\ref{tab2}. The $\dot{M}_{\star} v_{\infty}^{2}-M_{\star}$ relations displayed in Fig.~\ref{fig5} allow for the calculation of the total wind energy deposited by stellar winds. The total energy from the winds is given by:

\begin{equation}
E_{wind} = \int_{t'=0}^{t'=t} \int_{M_{\star}=5~\rm{M_{\odot}}}^{M_{\star}=120~\rm{M_{\odot}}} \left( \frac{N(M_{\star}) \dot{M_{\star}} (M_{\star}) v_{\infty}^{2}}{2} dM_{\star}\right) dt'. 
\label{eq20}
\end{equation}

We assume that only a fraction of $E_{wind}$ will be transformed into systemic motions that will oppose gravity and participate in the evacuation of the bulk of the gas from the proto-cluster clump. The rest of the energy is assumed to be dissipated in wind-wind collisions or escape the wind bubble. The effective kinetic wind energy is thus given by: 

\begin{equation}
 E_{k,wind}=\kappa~E_{wind},
\label{eq21}
\end{equation}
 
\noindent where $\kappa$ is a quantity $\leq 1$. It is currently difficult to estimate $\kappa$ as its exact value will vary from system to system depending on the number of massive stars, their locations, and their wind interactions. As a conservative guess for the fiducial model, we take $\kappa=0.1$. $E_{k,wind}$ is compared at every timestep to the absolute value of the gravitational energy, $E_{grav}$, which is calculated as being:

\begin{equation}
E_{grav} =  -\frac{16}{3} \pi^{2} G \int_{0}^{R_{c}} \rho_{c}(r)^{2} r^{4}  dr,
\label{eq22}
\end{equation}

\noindent where $\rho_{c}$ is given by Eq.~\ref{eq7}.

\section{RESULTS}\label{results}

\subsection{THE CO-EVOLUTION OF THE CORE MASS FUNCTION AND OF THE IMF: A FIDUCIAL MODEL}\label{co_evol_fiducial}

\begin{figure*}
\begin{center}
\includegraphics[height=18.5cm, width=15.5cm]{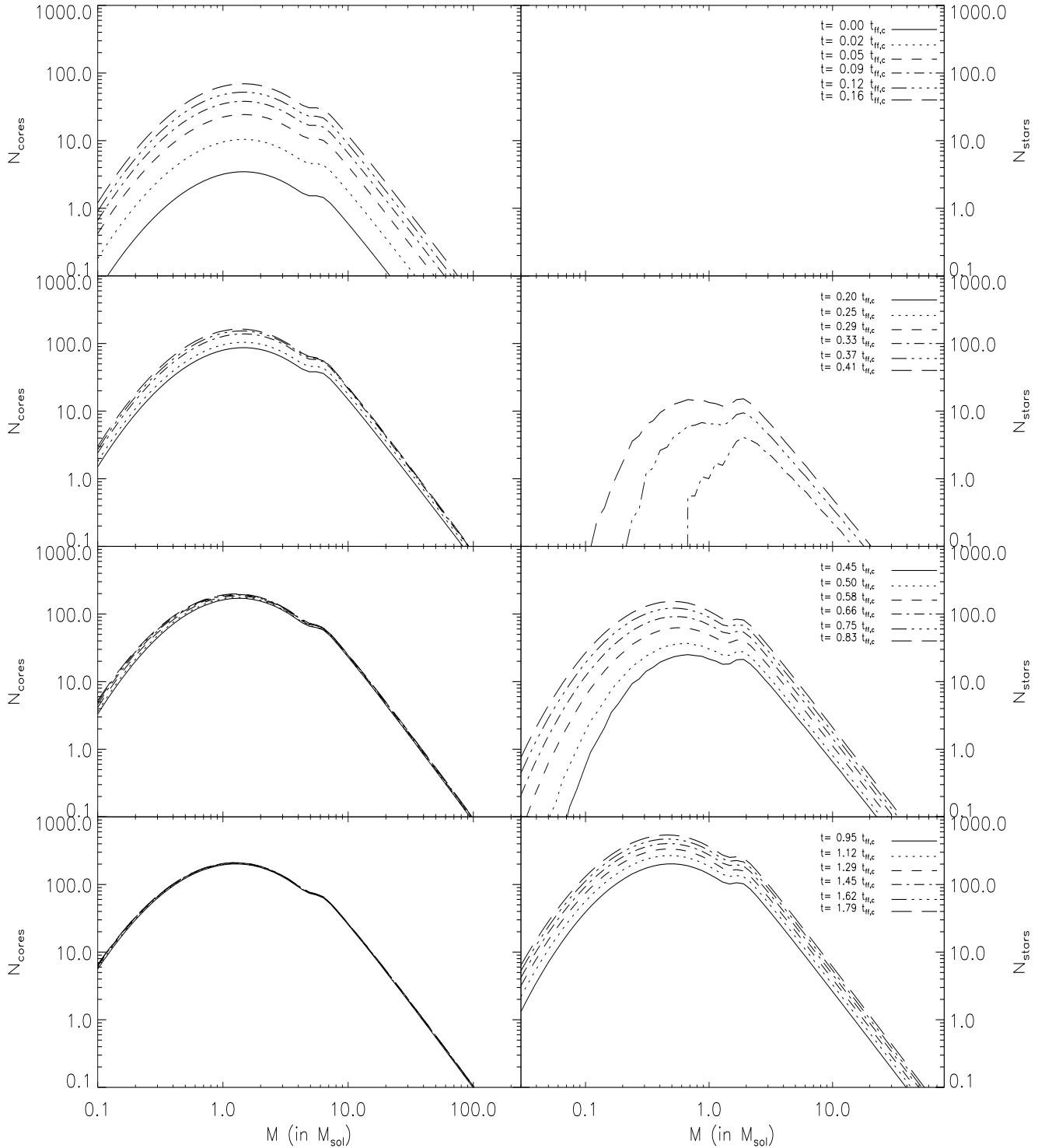}
\end{center}
\vspace{0.7cm}
\caption{Time evolution of the pre-stellar core mass function (left), and stellar mass function (right) in the protocluster clump with the fiducial model parameters. The last  time-step shown is $1.79~t_{ff,c}$ which corresponds to the epoch at which gas is expelled from the protocluster clump.}
\label{fig6}
\end{figure*}

Whenever a population of cores of mass $M$, located at a distance $r$ from the centre of the clump has evolved for a time that is equal to its contraction timescale, it is collapsed into stars. Thus, the local number of cores of a given mass, at a given epoch $\tau$, is the sum of all the local populations of cores of the same mass that have formed at all epochs anterior or equal to the considered epoch with the additional step of subtracting from that sum the cores of the same mass that have readily collapsed into stars. The local populations of  cores of various ages are evolved separately as they are each in a different phase of their contraction (and if accretion onto the cores were included as in Dib et al. 2010b, they will have different accretion histories and rates), and will collapse and form stars at various epochs. Thus, the total local number of cores of a given mass $M$, at a time $t$, is given by:

\begin{figure}
\begin{center}
\epsfig{figure=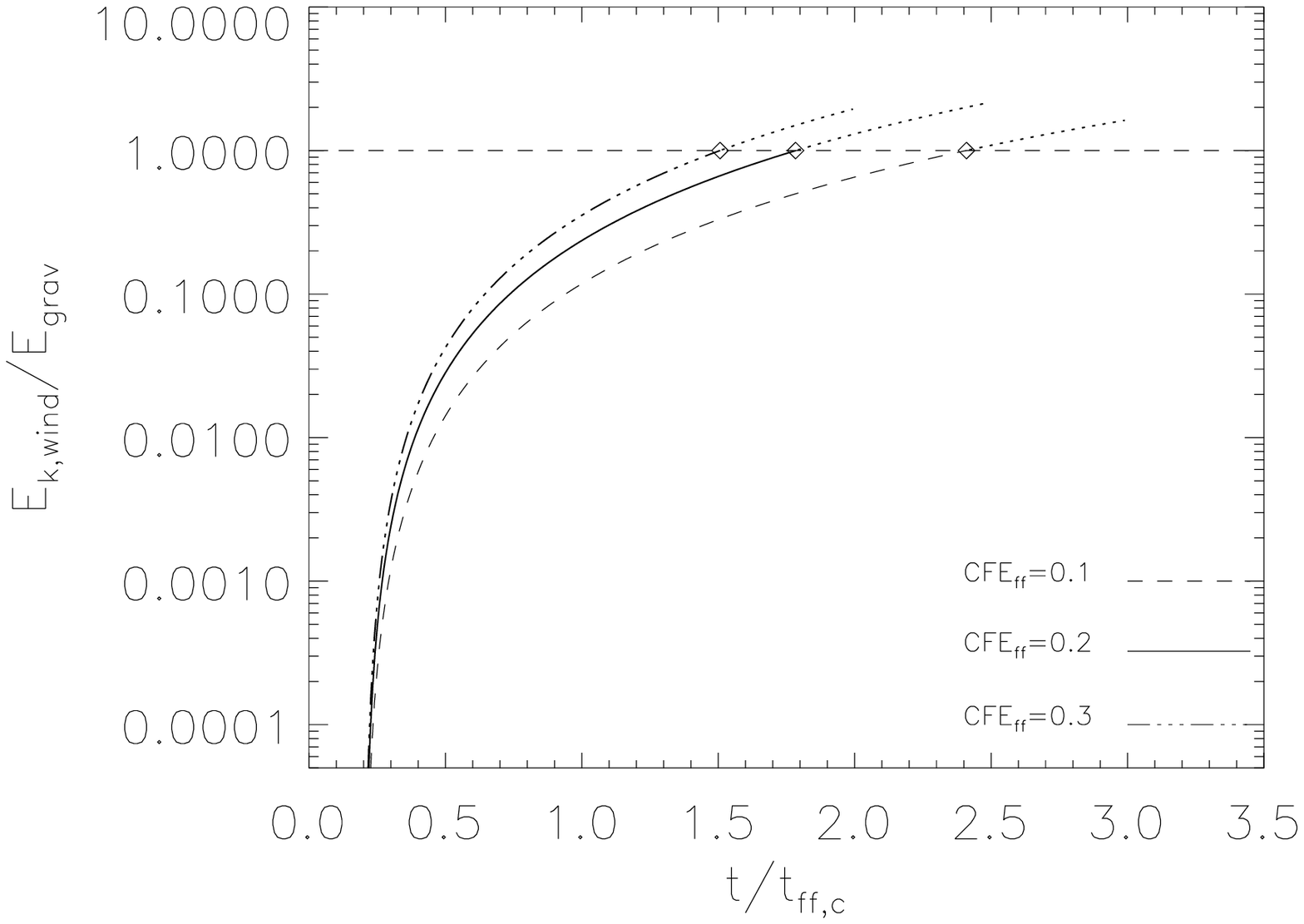,width=\columnwidth}
\end{center}
\caption{Time evolution of the ratio of the effective kinetic energy generated by stellar winds and the gravitational energy of the protocluster clump. Time is shown in units of the protocluster clump free fall timescale $t_{ff,c}$. The horizontal dashed corresponds to $E_{k,wind}/E_{grav} =1$ with  $\kappa =0.1$. The full line correspond to the fiducial model with the $CFE_{ff}=0.2$ and the dashed and triple dot-dashed to cases with $CFE_{ff}=0.1$ and $CFE_{ff}=0.3$, respectively.  Diamonds correspond to the epochs at which the gas is evacuated from the cluster in the three models and the processes of core and star formation are terminated. }
\label{fig7}
\end{figure}

\begin{figure}
\begin{center}
\epsfig{figure=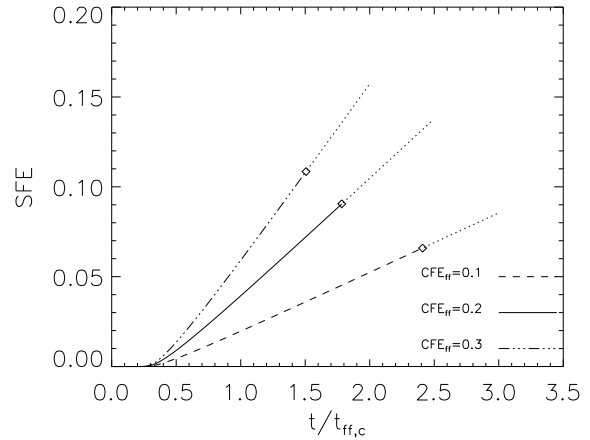,width=\columnwidth}
\end{center}
\caption{Time evolution of the SFE in the protocluster clump. The full line corresponds to the fiducial model with the $CFE_{ff}=0.2$ and the dashed and triple dot-dashed to cases with $CFE_{ff}=0.1$ and $CFE_{ff}=0.3$, respectively. Time is shown in units of the protocluster clump free fall timescale $t_{ff,c}$. Diamonds correspond to the epochs at which the gas is evacuated from the cluster and indicates the final value of the SFE, $SFE_{f}$, in the three models.}
\label{fig8}
\end{figure}

\begin{equation} 
N(r,M,t)=\sum_{\tau_{i}  \leq t} N(r,M,\tau_{i},t).
\label{eq23}
\end{equation}

We assume that only a fraction of the mass of a core ends up locked in the star. We account for this mass loss in a purely phenomenological way by assuming that the mass of a star which is formed out of a core of mass $M$ is given by M$_{\star}=\xi M$, where $\xi \le 1$. Matzner \& McKee showed that $\xi$ can vary in the range $0.25-0.75$. In the absence of strong observational or theoretical constraints, we shall assume that $\xi$ is independent of the core mass. Finally, we account for the possible modification of the stellar mass function by the effect of stellar winds at the high mass end. The variations in the IMF at the high mass end will be given by:

\begin{eqnarray} 
\left(\frac{dN_{\star}(r,M,t)}{dt}\right)= \nonumber \\
\left[\left(\frac{\partial N_{\star}}{\partial M_{\star}} \right) \dot{M_{\star}}+\left(\frac{\partial \dot{M_{\star}}}{\partial M_{\star}}\right) N_{\star}\right] (r,M,t),
\label{eq24}
\end{eqnarray}

\noindent where $\dot{M}_{\star}$ is the stellar mass loss rate given by Eqs.~\ref{eq15}-\ref{eq16}, and $N_{\star}$(r,M,t) is the local number of stars, at time $t$, of mass $M_{\star}$. 

In this section, we describe in detail the results for a fiducial model. In this model, the mass of the clump is taken to be M$_{clump}=10^{5}$ M$_{\odot}$, the metallicity is $Z=Z_{\odot}$, and The CFE is $CFE_{ff}=0.2$, where $t_{ff,cl}$ is the free-fall time of the clump and is given by $t_{ff,cl}= (3 \pi/ 32~G \bar{\rho_{c}})^{1/2}$, and $\bar{\rho_{c}}$ the average density of the clump. The other quantities have been taken to be equal to the most commonly cited observational determinations and have been set, in the fiducial model as well as in all other  models, to the following values: $R_{c0}=0.2$ pc, $\nu=3$, $\mu=0.2$, $\kappa=0.1$. We also assume that $\xi=1/3$ which implies that 1/3 of the mass of the dense core ends up in the final star mass. It is worth mentioning here that for clump masses of the order of a few $10^{5}$ M$_{\odot}$ or less and in which stars form with a global SFE of a few up to a few tens of percents, the role of direct radiation pressure will be negligible and stellar winds are likely to be the dominant mechanism for the expulsion of the gas from the protocluster region. In this case, as the luminosities of OB stars are almost independent of metallicity (i.e., Fig.~\ref{fig3}), any role played by direct radiation pressure will be to accelerate the expulsion of the gas from the protocluster region without modifying the dependence of the stellar winds on metallicity. 

Fig.~\ref{fig6} displays the time evolution of the CMF (left column) and of the IMF (right column) in the fiducial model. Each population of cores formed at a given epoch is the sum of all populations of cores formed at different positions in the protocluster clump. The feature that appears at $\sim 6$ M$_{\odot}$ is a result of this summation. By $t \sim 0.3~t_{ff,cl}$, the first stars form. In this model, since $\mu=0.2>~ 0$, the most massive stars form first, as the most massive cores tend to be, at any given position in the clump, more centrally peaked and thus have shorter lifetimes. As time advances, the IMF becomes fully populated. On the other hand, the CMF ceases to evolve (i.e., there is no accretion or coalescence in this model as in Dib 2007, Dib et al. 2007b; Dib et al. 2008b; and Dib et al. 2010a) as the numbers of cores that are newly formed at each position in the protocluster clump is balanced by an equal number of cores which collapses and forms stars. The final IMF of the cluster is established at $\sim 1.79~t_{ff,c}$, which corresponds to the last epoch shown in Fig.~\ref{fig6}. At this epoch, Fig.~\ref{fig7} shows that the ratio $E_{k,wind}/E_{grav}$ reaches unity (full line), and as a consequence, gas is expelled from the protocluster region. Fig.~\ref{fig8} displays the time evolution of the SFE in this model (full line). At $t=t_{exp}=1.79~t_{ff,cl}$, the final value of the SFE is $SFE_{f} \sim 9.05 \times 10^{-2}$.
   
In this fiducial model, we have adopted a CFE value of $CFE_{ff}=0.2$. This is an intermediate value between the values of $\sim 0.06$ and $\sim 0.33$ measured by Dib et al. (2010a) in numerical simulations of molecular clouds with two different degree of magnetization (moderately and strongly super-critical clouds, respectively). At first glance, it may appear that increasing the CFE by a given factor will lead to an increase in $SFE_{f}$ by approximately the same factor. However, for a fully sampled IMF, a larger CFE value implies that a larger number of OB stars will be formed and deposit larger amounts of feedback by stellar winds in the protocluster region. This in turn leads to a faster evacuation of the gas and to a limitation of the $SFE_{f}$. Fig.~\ref{fig7} shows that in two other models similar to the fiducial case but with a different CFE, the evacuation of the gas occurs faster for the higher CFE case (i.e., case with $CFE_{ff}=0.3$) and slower for the low CFE case (i.e., $CFE_{ff}=0.1$). The time evolution of the SFE in these two additional models is compared to the fiducial case in Fig.~\ref{fig8}. The final values of the SFE at $t=t_{exp}$ are $SFE_{f} \sim 6.59 \times 10^{-2}$, $\sim 9.05 \times 10^{-2}$, and $\sim 0.11$ for the cases with $CFE_{ff}=0.1, 0.2,$ and $0.3$ respectively. A factor of 2 and 3 increase in the CFE leads to an increase of the $SFE_{f}$ only by factors of $\sim 1.37$ and $\sim 1.64$, respectively. This implies a strong regulation of the effect of a varying CFE by stellar feedback on the resulting $SFE_{f}$ in protocluster clumps.      

Another quantity that can modify the value of the $SFE_{f}$ is the mass-radius scaling relation of the protocluster clump. In this section and in the following ones, we have used the mass-radius relation of Saito et al. (2007) which is based on C$^{18}$O (1-0) line observations of massive clumps. With this relation, the radius of a $10^{5}$ M$_{\odot}$ clump is $\sim 3.5$ pc (Eq.~\ref{eq5}). Any other putative mass-radius relation that would for example result in the decrease of the radius of the clump $R_{c}$ by a factor of $2$ (for the same mass and clump core-radius values), would lead to the reduction of the gravitational energy by a factor of $\sim 4$ (i.e., $\rho_{c0}$ increase by a factor of 2, Eq.~\ref{eq8}) and an increase in the effective wind energy by a factor of $\sim 16$ (enhancement in the number of OB stars by a factor of 2, and the feedback energy dependence is $\propto M_{OB}^{4}$, where $M_{OB}$ is the total mass of OB stars). In this case, the gas will be expelled earlier and the value of $SFE_{f}$ will be smaller. There are however some indirect indications that the Saito et al. (2007) relations provide a good description of the scaling relations that connect the masse, radii, and velocity dispersions of massive protocluster clumps. As already discussed by Dib et al. (2010b), the bulk of the star formation in the clump occurs in the inner regions (between $3-5~R_{c0}$) (see Fig.~8 in Dib et al. 2010b). With the adopted value of $R_{c0}=0.2$ pc, the bulk of the stellar mass will form in a region of up to $\sim 1$ pc radius. Gas expulsion is conducive to cluster expansion (e.g., Bastian \& Goodwin 2006). Cluster expansion after gas removal will be visible in the form of a deviation from the initial density profile in the outer regions of the cluster and would increase the observed cluster radius by a factor of ~ 2-2.5 (i.e., for a cluster mass of $M_{cl}=7.5 \times 10^{4}$ M$_{\odot}$ after $20$ Myrs and by a similar factor for a $5 \times 10^{4}$ M$_{\odot}$ after $40$ Myrs ). This would lead the expanding clusters formed from clumps that follow the Saito et al. (2007) scaling relations to have radii $\sim 2-2.5$ pc. This agrees very well with the observed mean value of the stellar clusters radii in M51 obtained by Scheepmaker et al. (2007). 

\subsection{THE METALLICITY-STAR FORMATION EFFICIENCY RELATION}\label{metal_dep}

\begin{figure}
\begin{center}
\epsfig{figure=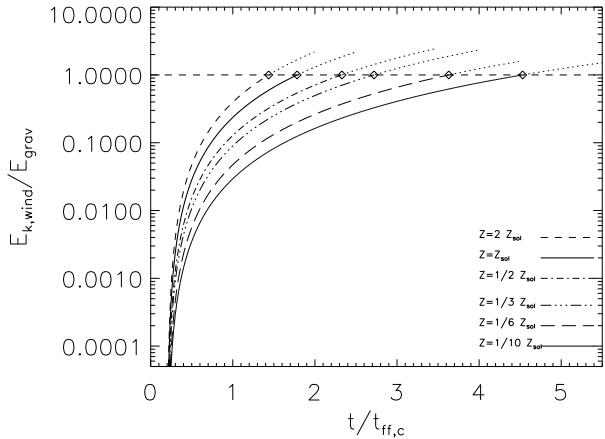,width=\columnwidth}
\end{center}
\caption{Time evolution of the ratio of the kinetic energy generated by stellar winds and the gravitational energy of the protocluster clump. Time is shown in units of the protocluster clump free fall timescale $t_{ff,c}$. The horizontal dashed corresponds to $E_{k,wind}/E_{grav} =1$ with $\kappa=0.1$. Diamonds correspond to the epochs at which the gas is evacuated from the cluster in the three models. The mass of the clump in these models is $10^{5}$ M$_{\odot}$.}
\label{fig9}
\end{figure}

\begin{figure}
\begin{center}
\epsfig{figure=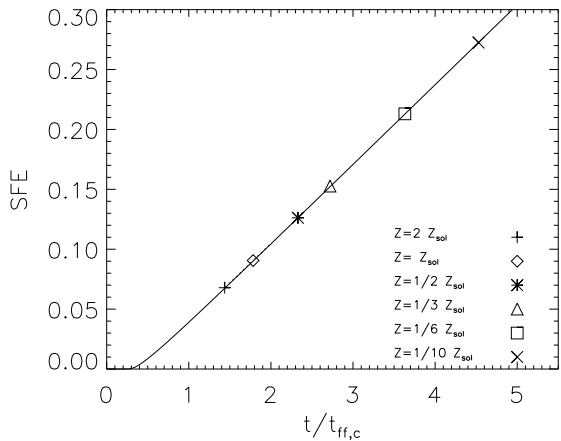,width=\columnwidth}
\end{center}
\caption{Time evolution of the SFE in the protocluster clump. Time is shown in units of the protocluster clump free fall timescale $t_{ff,c}$. The symbols correspond to the epochs at which the gas is evacuated from the cluster and indicates the final value of the SFE , $SFE_{f}$, for the models with different metallicities.}
\label{fig10}
\end{figure}

\begin{figure}
\begin{center}
\epsfig{figure=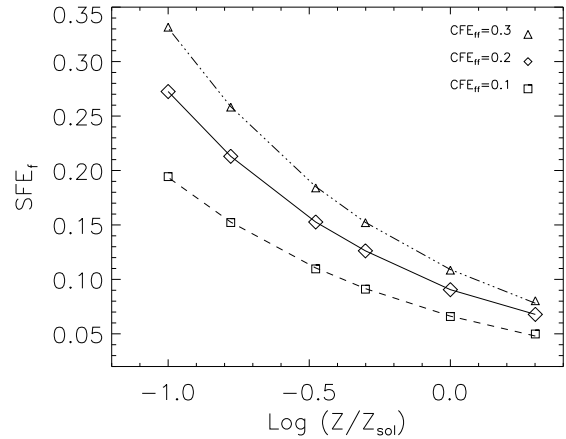,width=\columnwidth}
\end{center}
\caption{$SFE_{f}$-Metallicity relations for clumps with the fiducial mass of $10^{5}$ M$_{\odot}$ and for different core formation efficiencies of $CFE_{ff}=0.1$, $0.2$, and $0.3$. Over-plotted to the data are fit functions (Eq.~\ref{eq25}) whose parameters are listed in Tab.~\ref{tab4}.}
\label{fig11}
\end{figure}

We now turn to the effect of varying the metallicity in the protocluster clump. Fig.~\ref{fig9} displays the time evolution of the ratio $E_{k,wind}/E_{grav}$ in models similar to the fiducial case but with metallicities varying between $Z=0.1~Z_{\odot}$ and $Z=2~Z_{\odot}$. All six models have the fiducial value of the CFE, $CFE_{ff}=0.2$. In models with lower metallicities, the power of the stellar winds in weaker (i.e., Fig.~\ref{fig5}) and the evacuation of the gas occurs at later epochs as compared to the higher metallicity cases. In the model with $Z=2~Z_{\odot}$, the gas is expelled from the protocluster region at $t \sim 1.4~t_{ff,cl}$ while in the model with $Z=0.1~Z_{\odot}$, the gas expulsion is delayed until $t \sim 4.5~t_{ff,cl}$. For a given $CFE$, longer timescales imply a larger $SFE_{f}$. Fig.~\ref{fig10} displays the time evolution of the SFE in all six models. The curve is similar for all six models as the structural and dynamical properties of the models are similar (only the metallicty is different). The different symbols on the curve in Fig.~\ref{fig10} mark the epochs at which gas is expelled from the protocluster clump for the different metallicity cases and indicate the corresponding values of the $SFE_{f}$ in each model. These values are plotted versus metallicity in Fig.~\ref{fig11}. A clear trend is observed in which the $SFE_{f}$ increases with decreasing metallicity (diamonds). We fit the $SFE_{f}-Z$ points with the following functional form:

 \begin{equation} 
SFE_{f}= C_{Z}~e^{-\frac{1}{\tau_{Z}} \log \left(\frac{Z}{Z_{\odot}} \right)},
\label{eq25}
\end{equation}
 
\noindent with $C_{Z}=0.091 \pm 6.8\times 10^{-4}$ and $\tau_{Z}=0.91\pm 7.9 \times 10^{-3}$. We have also repeated the calculations at the various metallicities using the additional values of $CFE_{ff}=0.1$ and $CFE_{ff}=0.3$. The dependence of the $SFE$ on metallicity in these cases are also shown in Fig.~\ref{fig11}. As in the solar metallicity case, Fig.~\ref{fig11} shows that the $SFE_{f}$ vary typically by a factor of $\sim 1.3-1.4$ and $\sim 1.6-1.7$ for variations of the CFE by factors of 2 and 3, respectively. The curves corresponding to the cases with $CFE_{ff}=0.1$ and $CFE_{ff}=0.3$ were also fitted with the functional form given in Eq.~\ref{eq25}. The values of the fit parameters are summarized in Tab.~\ref{tab4}.      

\begin{figure}
\begin{center}
\epsfig{figure=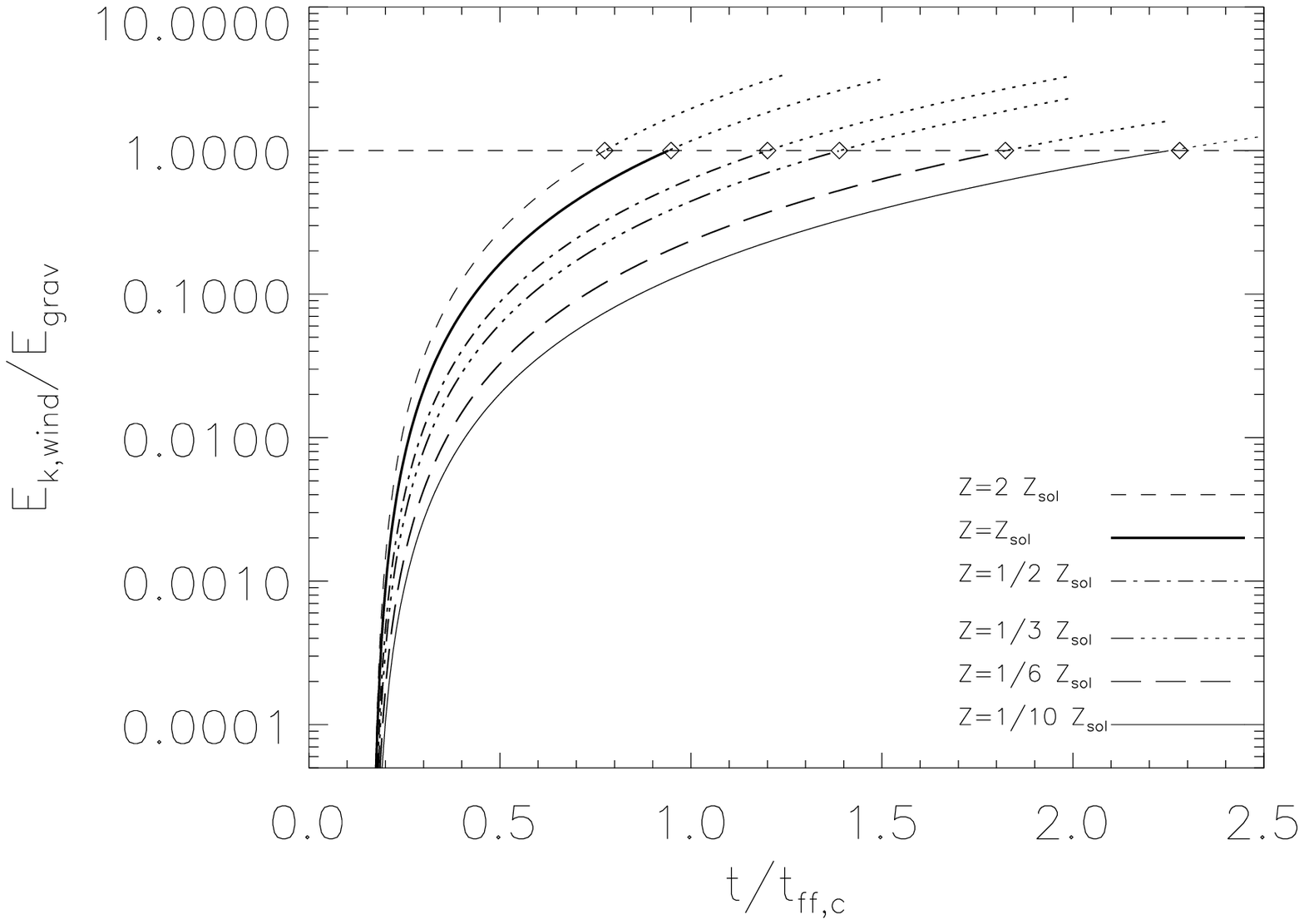,width=\columnwidth}
\end{center}
\caption{Time evolution of the ratio of the effective kinetic energy generated by stellar winds and the gravitational energy of the protocluster clump. Time is shown in units of the protocluster clump free fall timescale $t_{ff,c}$. The horizontal dashed corresponds to $E_{k,wind}/E_{grav} =1$ with $\kappa=0.1$. Diamonds correspond to the epochs at which the gas is evacuated from the cluster in the three models. The mass of the clump in these models is $5\times 10^{4}$ M$_{\odot}$.}
\label{fig12}
\end{figure}

\begin{figure}
\begin{center}
\epsfig{figure=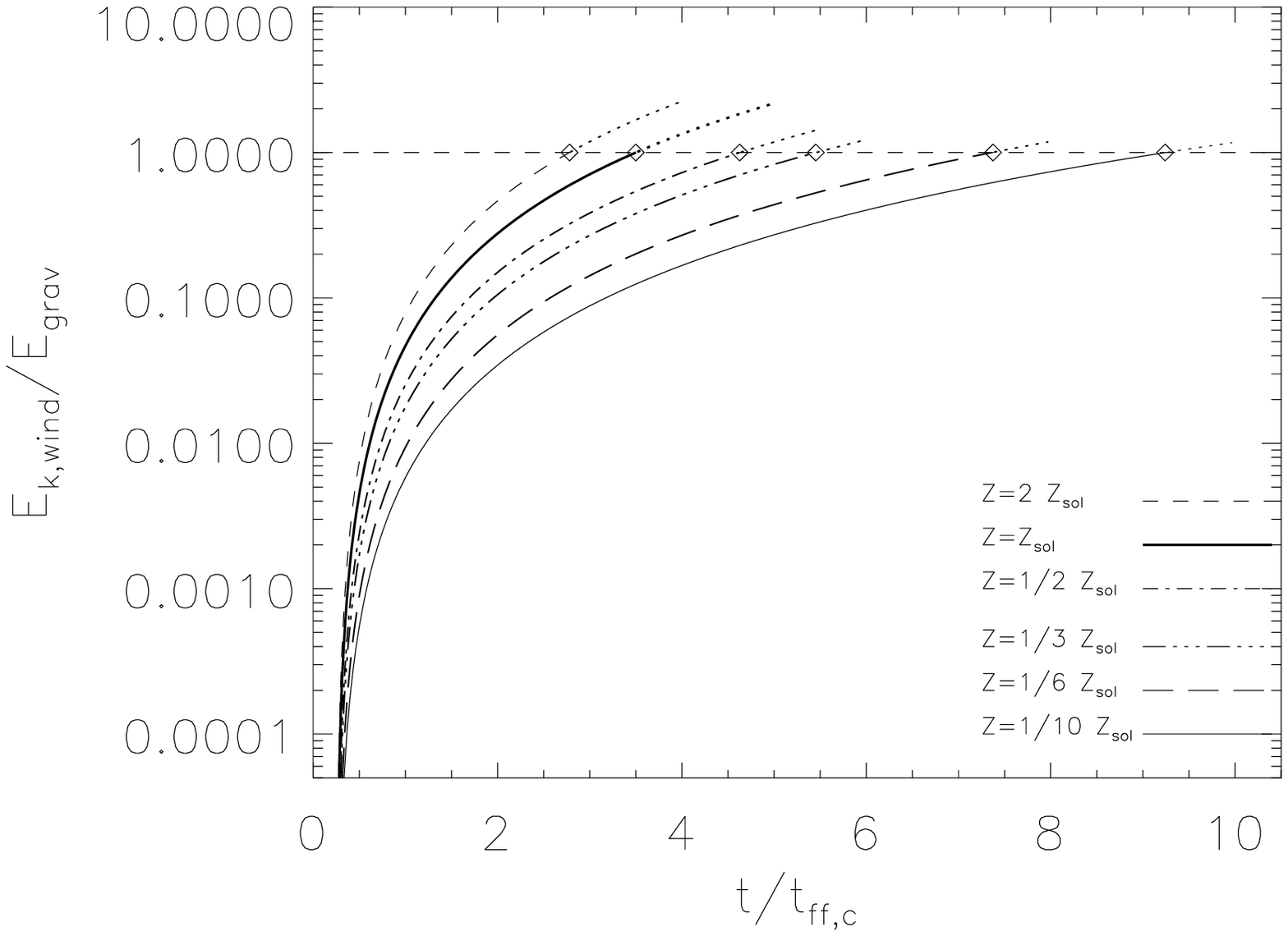,width=\columnwidth}
\end{center}
\caption{Time evolution of the ratio of the effective kinetic energy generated by stellar winds and the gravitational energy of the protocluster clump. Time is shown in units of the protocluster clump free fall timescale $t_{ff,c}$. The horizontal dashed corresponds to $E_{k,wind}/E_{grav} =1$ with $\kappa =0.1$. Diamonds correspond to the epochs at which the gas is evacuated from the cluster in the three models. The mass of the clump in these models is $2 \times10^{5}$ M$_{\odot}$.}
\label{fig13}
\end{figure}

\begin{figure}
\begin{center}
\epsfig{figure=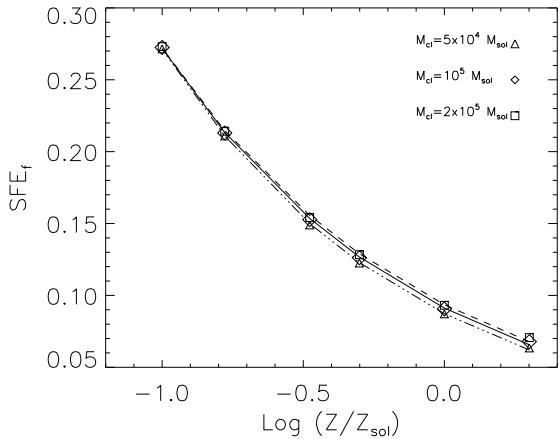,width=\columnwidth}
\end{center}
\caption{$SFE_{f}$-Metallicity relations for clumps of different masses with the fiducial core formation efficiency of $CFE_{ff}=0.2$. Over-plotted to the data are fit functions (Eq.~\ref{eq25}) whose parameters are listed in Tab.~\ref{tab5}.}
\label{fig14}
\end{figure} 

\subsection{THE DEPENDENCE OF THE SFE-Z RELATION ON THE MASS OF THE PROTOCLUSTER CLUMP}\label{mass_dep}

In the previous section, we have explored the dependence of the SFE on metallicity for a fiducial clump mass of $10^{5}$ M$_{\odot}$. For a given value of the $CFE_{ff}$, If the mass of the clump decreases (increases), the number of stars including OB stars decreases (increases) and the feedback from stellar winds decreases (increases). Simultaneously however, when the clump mass decreases (increases), the gravitational energy of the clump decreases (increases) and this allows for a shorter (longer) delay before gas expulsion from the protocluster clump starts hence for the buildup of a smaller (higher) SFE. In order to explore the effects on the $SFE_{f}-Z$ relation of these two opposing dependencies, we have performed additional models with clump masses increased and reduced by a factor of 2 from the fiducial clump mass value. Fig.~\ref{fig12} displays the time evolution of the ratio $E_{k,wind}/E_{grav}$ in the model with $M_{cl}=5\times 10^{4}$ M$_{\odot}$. In comparison to cases with the fiducial mass (i.e., Fig.~\ref{fig9}), the evacuation of the gas from the protocluster region occurs on shorter timescales. In contrast, Fig.~\ref{fig13} shows that the evacuation of the gas from a clump with a mass twice as large as the fiducial mass (i.e., $2 \times 10^{5}$ M$_{\odot}$) occurs on timescales (in units of the corresponding free-fall timescale of the clump) larger than those with the fiducial mass for the different metallicity cases. 

Fig.~\ref{fig14} compares the $SFE_{f}-Z$ relation for the models with the three clump masses. The effect of reducing (increasing) the amount of feedback from winds by decreasing (increasing) the clump mass is compensated by the effect of the reduction (increase) of the gravitational potential. This results in final SFE values at the different metallicities that are nearly identical and independent of the clump mass. The fit parameters for the data displayed in Fig.~\ref{fig14} are summarized in Tab.~\ref{tab5}. This negligible dependence of the $SFE_{f}-Z$ relation on clump mass is expected to remain valid for stellar clusters originating from clumps whose masses are smaller than the ones considered here, at least when averaged over a large number of clumps. On the other hand, the case of more massive clumps which would result in stellar clusters whose masses are larger than those considered here would require taking into account the effect of direct radiation pressure (e.g., Krumholz \& Matzner 2009). This is left to a subsequent work. Note however that very massive clusters are only found in large numbers in starburst galaxies such as the Antennae or M82. 

The range of $SFE_{f}$ values we obtained as a function of the adopted ranges in metallicity and $CFE_{ff}$ is $0.05-0.35$. A number of studies which have measured the fraction of bound stars and cluster survival following an instantaneous gas expulsion of the gas from the protocluster region indicate that a minimum value of $SFE_{f} \sim {0.33}$ is necessary for cluster survival (Lada et al. 1984; Geyer \& Burkert 2001, Boily \& Kroupa 2003a,b; Fellhauer \& Kroupa 2005; Baumgardt \& Kroupa 2007; see summary in Fig. 1 of Parmentier \& Gilmore 2007)\footnote{One way of increasing the final SFE, is by increasing the $CFE_{ff}$. We have adopted realistic values of the $CFE_{ff}$ in the range $0.1-0.3$. Adopting $CFE_{ff}=0.5$ will shift the SFEs obtained at the different metallicities by a factor of $\sim 1.5$ from their current values with the $CFE_{ff}$ of $0.2-0.3$. However, the $SFE_{f}$ for the solar metallicity case will still remain close to $\sim 0.15$}. It is important to note that these simulations did not explicitly model the dynamics of the gas. Instead, gas expulsion was performed by a reduction of the cluster potential which mimics a gas depletion process over a given timescale. For the range of SFEs found in our work, only models in which gas was removed on timescales of the order of 10 times the dynamical timescale of the cluster resulted in bound clusters. In these latter models, stars were assigned initial velocity dispersions such that the entire system (gas+stars) is in virial equilibrium. This most likely contradicts the results of many gravo-turbulent star formation simulations in which stars form in dense and quiescent regions of the clouds/clumps and in which most of the turbulent motions have been dissipated (e.g., Klessen et al. 2000, Dib et al. 2007a). In the framework of the theory of gravo-turbulent fragmentation of a cloud/clump, which we envision to be responsible for core formation in the protocluster clump considered here, stars will inherit the low levels of velocity dispersions from their gaseous progenitors. Geyer \& Burkert (2001) performed a direct  Nbody+SPH simulations in order to model the behaviour of stars and gas more consistently while requiring that stars form with zero velocity dispersion. They found that this configuration leads to the formation of compact clusters and to a severe reduction of the SFE required to form bound clusters (50 percent of bound stars for a SFE of 0.05 and close to 70 percent for a SFE of 0.1; Fig.~7 in their paper). Further work is obviously needed to further clarify the dependence of the stellar cluster survivability on the SFE using models that implement more realistic gas expulsion schemes.               

\section{IMPLICATIONS FOR  THE EMBEDDED CLUSTER MASS FUNCTION}\label{m_m_relation}

\begin{figure}
\begin{center}
\epsfig{figure=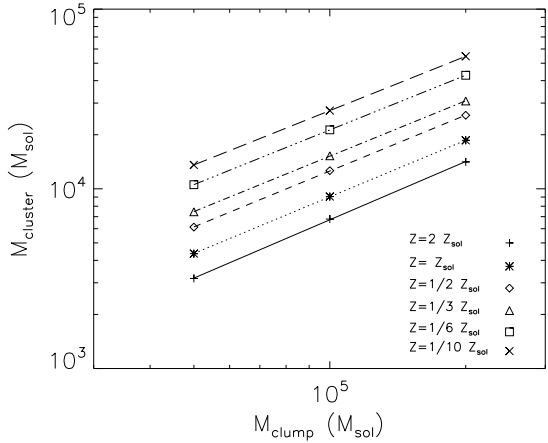,width=\columnwidth}
\end{center}
\caption{Clump mass-embedded stellar cluster mass relation in models of different metallicities and for the fiducial CFE value of $CFE_{ff}=0.2$.}
\label{fig15}
\end{figure} 
  
An interesting application of the results presented above is to explore the relationship between the exponent of the protocluster clumps mass function and that of the mass function of embedded clusters. Furthermore, it is interesting to explore to which extent this relation depends on metallicity. Fig.~\ref{fig15} displays the stellar clusters masses as a function of their parental gas clump masses for the various metallicities considered in this work. All cases presented here are for the fiducial CFE value of $CFE_{ff}=0.2$. The $M_{clump}-M_{cluster}$ relations in Fig.~\ref{fig15} can be approximated by power laws of the form $M_{cluster}=C_{m}~M_{clump}^{\gamma}$. If the clump mass function follows a power law of the form $dN/dM_{clump} \propto M_{clump}^{\delta}$ and the young clusters mass function another power law given by $dN/dM_{cluster} \propto M_{cluster}^{\eta}$, then the exponents $\gamma$, $\delta$, and $\eta$ are bound by the relation $\delta=\gamma~(\eta +1)-1$. Tab.~\ref{tab6} summarizes the values of $\gamma$ for the $M_{clump}-M_{cluster}$ relations for the various metallicity cases. The derived values of $\gamma$, which are close to unity, suggest that there should exist only a small difference between the exponent of all the mass functions of protocluster clumps and that of the young stellar clusters. In fact, if $\eta$ falls in the range $-1.80$ to $-2$ as is observed for young stellar clusters in the Galaxy and in other spiral galaxies (e.g., Elmegreen \& Efremov 1997; McKee \& Williams 1997; Zhang \& Fall 1999; Dowell et al. 2008; Larsen 2009; Chandar et al. 2010), then the derived values of $\delta$ fall in the range $[-1.753,-1.808]$ (for $\eta=-1.75$) and $[-2.004,-2.077]$ (for $\eta=-2$), for metallicities in the range of $Z/Z_{\odot}=[0.1,2]$, respectively. 

In a recent paper, Fall et al. (2010) suggested that a momentum driven feedback, of any kind, is sufficient to relate the exponent of the young clusters mass function ($\eta \sim -2$) and the value of $\delta \sim -1.7$ that is quoted by the latter authors as being the exponent of the protocluster clumps mass function. In reality, there is little evidence that the exponent of $\delta \sim -1.7$ relates to the mass functions of gravitationally {\it bound} structures. Dib et al. (2008b) showed that the mass function of gravitationally unbound, turbulence dominated, cores/clumps is well described by a power law whose exponent is $\sim -1.7$. A similar result has been obtained by Hennebelle \& Chabrier (2008). Dib et al. (2008a) also found that the mass function steepens with the increasing density threshold and therefore increasing level of the gravitational boundedness of the selected cores/clumps, reaching values of $\gtrsim -2$ for populations of cores/clumps which are entirely gravitationally bound (i.e., this happens when cores/clumps are detected at threshold densities of $\gtrsim 10^{5}$ cm$^{-3}$, Dib et al. 2007a,, Dib \& Kim 2007). Using a similar argument based on the existence of a threshold density for star formation to occur in the dense regions of lower density molecular clouds/clumps (as argued for by Dib et al. 2007a), Parmentier (2011) showed that the explains the difference between the slope of the mass function of low density clouds/clumps ($\sim -1.7$) and that of high density protocluster forming clumps ($\sim -2$).     
  
\section{IMPLICATIONS FOR GALAXIES AND GLOBULAR CLUSTERS}\label{impli_galaxies}

\begin{figure}
\begin{center}
\epsfig{figure=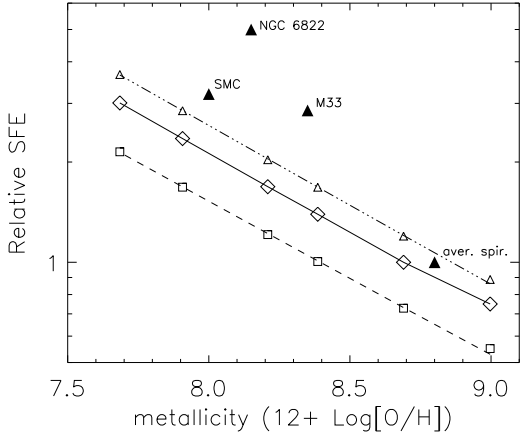,width=\columnwidth}
\end{center}
\caption{Relative $SFE$ as a function of metallicity. The relative $SFE$ for the models is taken to be $SFE_{ff}/SFE_{f}(Z=Z_{\odot},CFE_{ff,}=0.2)$. The relative $SFE$ for the observational data is $SFE_{gal}/SFE_{gal} (aver. spir.)$. The full line, dashed line, and triple-dot dash lines corresponds to the model with $CFE_{ff}=0.2$, $0.1$, and $0.3$, respectively.}
\label{fig16}
\end{figure}

There is, to date, no known observationally derived $SFE_{f}-Z$ relation for individual stellar clusters that can be directly compared to our models. It is however possible to establish such relation on galactic scales. Fig.~\ref{fig1} clearly shows that the galactic SFE, $SFE_{gal}$, decreases with increasing metallicity of the gas. The negligible dependence we find of the $SFE_{f}$ on the clump mass suggests that this $SFE_{f}-Z$ relation remains valid on galactic scales. This is true as long as star formation in clusters is the main culprit in establishing this relation and that other processes such as gas infall onto the galactic disk and galactic outflows play only a minor role. One way of comparing our $SFE_{f}-Z$ relation(s) to the galactic  $SFE_{gal}-Z$ relations is to calculate the relative SFE variations with respect to a characteristic value. For our stellar cluster models, we chose this value to be the $SFE_{f}$ at solar metallicity and for fiducial CFE value of $CFE_{ff}=0.2$. For galaxies, we normalize the $SFE_{gal}$ by the 'average spirals'  value. Fig.~\ref{fig16} displays the relative SFE, $SFE_{rel}$ as a function of metallicity expressed in $12+ \log[\rm{O/H}]$. The normalized relation corresponding to the fiducial case with $CFE_{ff}=0.2$ is fitted with the functional form given by: 

 \begin{eqnarray} 
SFE_{rel}=\frac {SFE_{f}} {SFE_{f}(Z=Z_{\odot},CFE_{ff}=0.2)} \nonumber \\
                 = C_{Z,rel}~e^{-\frac{1}{\tau_{Z,rel}} \log \left(\frac{Z}{Z_{\odot}} \right)},
\label{eq26}
\end{eqnarray}

yielding $C_{Z,rel}=1.007 \pm 7.5\times 10^{-3}$ and $\tau_{Z,rel}=0.915 \pm 7.0 \times 10^{-3}$.  The normalized relation corresponding to the cases with $CFE_{ff}=0.1$ and $0.3$ yield fit coefficients and exponents which differ by about $\sim 0.3$ dex and 0.03 from the fiducial case, respectively. The detailed values of all fit parameters are reported in Tab.~\ref{tab7}.    
 
In this work, we have explored the dependence of the final SFE on metallicity in the metallicity range $Z/Z_{\odot}=[0.1,2]$. The power of stellar winds is expected to decrease further at metallicities lower than $Z/Z_{\odot}=0.1$. Thus one may expect that for halo like metallicities which are those of old globular cluster, the $SFE_{f}$ will be substantially higher. The globular clusters in the Galaxy and in nearby galaxies are observed to be more massive than 'standard' open clusters (e.g., Mandushev et al. 1991; Fan et al. 2010). One explanation of this is that globular clusters formed in an environment of high density and low angular momentum. They are also thought to have been more massive at birth but were affected by evaporation processes since the time of their formation. Our results which suggest, if extrapolated to halo like metallicities, that the $SFE_{f}$ in globular clusters parental clumps is very high further strengthens the idea that the initial masses of globular clusters at the moment of their formation must have been many times higher than their currently measured mass (e.g., Schaerer \& Charbonnel 2011). It should be noted however that many globular clusters show signature of multiple stellar populations (e.g., Bekki 2011 and refereces therein). The gaseous ejecta from the first generation of AGB stars can be converted into new stars and this will alter the original $SFE_{f}-Z$ relations. In a future work, we plan to investigate the $SFE_{f}$ in clumps with halo like metallicities, paying particular attention to the effect of initial conditions (i.e., the shape and characteristic mass of the CMF). 
  
\section{SUMMARY}\label{summary}

We use semi-analytical modelling to study the effects of gas dispersal from a protocluster clump by stellar winds of various metallicities in the range $Z/Z_{\odot}=[0.1,2]$. In this model, populations of gravitationally bound cores form uniformly in time at different positions from the protocluster clump centre by turbulent fragmentation. The cores collapse to form stars after evolving over their lifetimes which are taken to be a few times their free fall times. The population of newly formed OB stars power strong stellar winds whose cumulative energy counters gravity and acts to disperse gas from the protocluster clump and hence quench further core and star formation. In order to estimate the power of the stellar winds, we use the stellar evolution code CESAM to calculate the basic stellar properties (i.e., effective temperature, stellar radius, and luminosity) of main sequence OB stars in the mass range $[5-80]$ M$_{\odot}$. These calculations were performed for the metallicities of $Z/Z_{\odot}=1/10, 1/6, 1/3, 1/2, 1$ and $2$. The latter quantities were then used in the stellar atmosphere model of Vink et al. in order to derive the mass loss rates and power of the winds. Our main results are the following:
  
 a) For a given core formation efficiency (CFE), expressed in units of the protocluster clump free-fall time $CFE_{ff}$, the final star formation efficiency, $SFE_{f}$, which is set at the epoch at which the gas is evacuated from the protocluster clump, decreases with increasing metallicity. With the fiducial value of $CFE_{ff}=0.2$, we find that $SFE_{f}=C_{Z}~e^{-\log (Z/Z_{\odot})/\tau_{Z}}$, with $C_{Z}=0.091 \pm 6.8\times 10^{-4}$ and $\tau_{Z}=0.91\pm 7.9 \times 10^{-3}$. This result is a direct consequence of the fact that lower metallicity stars have weaker winds than their higher metallicity counterparts. Weaker winds delay the evacuation of the gas from the protocluster clump and lead to higher values of the final star formation efficiency. With the fiducial value of the CFE, the $SFE_{f}$ varies between a few percent ($\sim 0.06$ for $Z= 2~Z_{\odot}$) to a few tens of percent ($\sim 0.27$ for $Z=0.1~Z_{\odot}$).     
 
b) For a given clump mass, the $SFE_{f}$, depends weakly on the CFE. Variations of the CFE by factors of 2 and 3 lead to variations of the $SFE_{f}$ by factor of $\sim 1.3$ and $\sim 1.6$, respectively. 
   
c) For a given CFE, the value of the $SFE_{f}$ depends very weakly on the mass of the protocluster clump. This is due to the fact that any increase in the clump mass which results in an increase in the number of OB stars, and hence feedback from stellar winds, is balanced by the increase in the gravitational potential of the clump which acts to delay gas evacuation. We find that varying the clump mass by a factor of 2 around the fiducial value of $10^{5}$ M$_{\odot}$ in the metallicity range we have explored changes the $SFE_{f}$ only by factors of a few percent. This indicates that if the star formation efficiency in clusters is the main driver of the galactic star formation efficiency-metallicity relation, then our result of $SFE_{f}=C_{f}~e^{-\log (Z/Z_{\odot})/\tau_{Z}}$, with $C_{Z} \sim 0.1$ and  $\tau_{Z} \sim 0.91$ remains valid on galactic scales. We show that, when the cluster and galactic SFEs are normalized to a characteristic value (i.e., the corresponding SFE at solar metallicity), they are nearly identical (Fig.~\ref{fig16}). 
 
d) We find that the exponent, $\gamma$ of the clump mass-stellar cluster mass relation, (i.e., $M_{clump} \propto M_{cluster}^{\gamma}$) is close to unity and increases only very weakly with decreasing metallicity. This implies that the exponent of the young star cluster mass function and that of the gravitationally bound, protocluster clumps mass function are nearly identical. 
 
e) As a by-product of this study, we provide fit functions of the power delivered by stellar winds from OB stars on the main sequence in the mass range $[5,80]$ M$_{\odot}$ and for various metallicities in the range $Z/Z_{\odot}=[0.1,2]$. These relations can be easily implemented in other analytical or numerical works studying the effects of feedback from stellar winds on the interstellar medium in galaxies of various metallicities.

\section*{Acknowledgments}

We are grateful to the referee for her/his constructive and useful comments and suggestions which helped us clarify several sections of this paper. We would like to thank Jorick Vink, S{\o}ren Larsen, and Andreas Burkert for useful discussions.  S. D. and S. M. acknowledge support from STFC grant ST/H00307X/1. 
 
\begin{table*}
\centering
\begin{tabular}{lccccc}
\hline
Metallicity & $A_{4}$ & $A_{3}$ & $A_{2}$ & $A_{1}$ & $A_{0}$  \\ 
 \hline
$Z=1/10~Z_{\odot}$   &  $-6.821\times 10^{-7}$ & $1.502\times 10^{-4}$ & $-1.230\times 10^{-2}$ & $4.688\times 10^{-1}$  & $-13.757$ \\
$Z=1/6~Z_{\odot}$     &  $-6.868\times 10^{-7}$ & $1.512\times 10^{-4}$ & $-1.237\times 10^{-2}$ & $4.715\times 10^{-1}$  & $-13.642$\\
$Z=1/3~Z_{\odot}$     &  $-6.947\times 10^{-7}$ & $1.530\times 10^{-4}$ & $-1.253\times 10^{-2}$ & $4.772\times 10^{-1}$  & $-13.516$\\
$Z=1/2~Z_{\odot}$     &  $-7.054\times 10^{-7}$ & $1.551\times 10^{-4}$ & $-1.267\times 10^{-2}$ & $4.816\times 10^{-1}$  & $-13.458$\\
$Z=1~Z_{\odot}$        &  $-7.098\times 10^{-7}$ & $1.566\times 10^{-4}$ & $-1.285\times 10^{-2}$ & $4.902\times 10^{-1}$  & $-13.406$\\
$Z=2~Z_{\odot}$        &  $-9.744\times 10^{-7}$ & $1.954\times 10^{-4}$ & $-1.486\times 10^{-2}$ & $5.364\times 10^{-1}$  & $-13.670$\\
\end{tabular}
\caption{Coefficients of the fourth order polynomial fit to the $M_{\star}$-${\rm log_{10}} (\dot{M}_{\star})$ relation (${\rm log} (\dot{M}_{\star})=\sum_{i=0,4} A_{i}~M_{\star}^{i}$, applied to the data in Fig.~\ref{fig4}). The stellar mass, $M_{\star}$, is in solar mass units, M$_{\odot}$, and the stellar mass loss rate,  ${\dot{M}}_{\star}$ is in  M$_{\odot}~$yr$^{-1}$.}
\label{tab1}
\end{table*}

\begin{table*}
\centering
\begin{tabular}{lccccc}
\hline
Metallicity & $B_{4}$ & $B_{3}$ & $B_{2}$ & $B_{1}$ & $B_{0}$  \\ 
 \hline
$Z=1/10~Z_{\odot}$   & $-6.398\times 10^{-7}$ & $1.416\times 10^{-4}$ & $-1.168\times 10^{-2}$ & $4.515\times 10^{-1}$ & $21.884$\\
$Z=1/6~Z_{\odot}$     & $-6.455\times 10^{-7}$ & $1.427\times 10^{-4}$ & $-1.177\times 10^{-2}$ & $4.549\times 10^{-1}$ & $22.055$\\
$Z=1/3~Z_{\odot}$     & $-6.524\times 10^{-7}$ & $1.443\times 10^{-4}$ & $-1.190\times 10^{-2}$ & $4.597\times 10^{-1}$ & $22.257$\\
$Z=1/2~Z_{\odot}$     & $-6.622\times 10^{-7}$ & $1.463\times 10^{-4}$ & $-1.205\times 10^{-2}$ & $4.642\times 10^{-1}$ & $22.359$\\
$Z=1~Z_{\odot}$        & $-6.666\times 10^{-7}$ & $1.478\times 10^{-4}$ & $-1.222\times 10^{-2}$ & $4.725\times 10^{-1}$ & $22.485$\\
$Z=2~Z_{\odot}$        & $-8.928\times 10^{-7}$ & $1.809\times 10^{-4}$ & $-1.394\times 10^{-2}$ & $5.130\times 10^{-1}$ & $22.322$\\
\end{tabular}
\caption{Coefficients of the fourth order polynomial fit to the $M_{\star}$-${\rm log_{10}} (\dot{M}_{\star}~v_{\infty}^{2})$ relation (${\rm log} (\dot{M}_{\star}~v_{\infty}^{2})=\sum_{i=0,4} B_{i}~M_{\star}^{i}$, applied to the data in Fig.~\ref{fig5}). The stellar mass, $M_{\star}$, is in solar mass units, M$_{\odot}$, and the power of the stellar winds (i.e., energy per unit time) in  ${\dot{M}}_{\star}$ is in J s$^{-1}$.}
\label{tab2}
\end{table*}

\begin{table*}
\centering
\begin{minipage}{15cm}
\begin{tabular}{l l}
\hline
Protocluster clump variables & Meaning of the variables \\
\hline
$M_{clump}$ & Mass of the clump \\
$R_{c}$   &  Radius of the clump \\
$R_{c0}$ & Core radius of the clump \\
$v_{c}$ & Velocity disperion of the gas in the clump\\
$\rho_{c0}$ & Ventral density of the clump \\
$E_{grav}$ & Gravitational energy of the protocluster clump \\
$\alpha$ & Exponent of the velocity dispersion-size relation in the clump\\
$\beta$ & Exponent of the kinetic energy power spectrum in the clump \\ 
$t_{ff,cl}$ & Free-fall timescale of the clump \\ 
$CFE_{ff}$ & Core formation efficiency per unit $t_{ff,cl}$ \\
$SFE_{f}$ & Final star formation efficiency in the clump. \\
\hline
Core variables & Meaning of the variables \\
\hline
 $M$ & Mass of the core \\
 $\rho_{p0}$ & Central density of the core \\
$\mu$& Exponent of the $\rho_{p0}-M$ relation of the cores \\
$R_{p0}$ & Core radius of the core  \\
$R_{p}$ & Radius of the core \\
 $t_{ff}$ & Free-fall timescale of the core \\
$t_{cont,p}$ & Contraction timescale of the core \\
$\nu$ & Ratio of the contraction timescale of the core to its free-fall time \\
\hline
Stellar and Feedback variables & Meaning of the variables \\ 
\hline
$M_{\star}$ & masses of stars \\
$\xi$ & fraction of the mass of the core that ends up in the stars\\
$\dot{M}_{\star}$ & Mass loss rates from stars \\ 
$\dot{M}_{\star}~v_{\infty}^{2}$ & Power of the stellar winds \\
$E_{wind}$ & time integrated wind energy from OB stars stars  \\
$\kappa$ & Fraction of the wind energies that opposes the total gas+stars gravity in the clump \\
$E_{k,wind}$ & Effective wind kinetic energy \\
\hline
 \end{tabular}
\end{minipage}
\caption{Main variables of the model. From top to bottom, the first panel describes the protocluster clump variables, the second panel the dense cores variables, the third panel  the stellar and stellar feedback variables.}
\label{tab3}
\end{table*}

\begin{table*}
\centering
\begin{tabular}{cccc}
\hline
 $CFE_{ff}$ & $0.1$ & $0.2$ & $0.3$  \\ 
 \hline
 $C_{Z}$ & $6.62 \times 10^{-2} \pm 6.13 \times 10^{-4}$ & $9.11\times 10^{-2} \pm 6.89 \times 10^{-4}$ & $1.09 \times 10^{-1} \pm 7.30 \times 10^{-4}$  \\
 $\tau_{Z}$ & $9.32 \times 10^{-1} \pm 1.01 \times 10^{-2}$ & $9.15 \times 10^{-1} \pm 7.97 \times 10^{-3}$ & $9.01 \times 10^{-1} \pm 6.90 \times 10^{-3}$\\
\end{tabular}
\caption{Coefficients and exponents of the relation between the final SFE of the cluster $SFE_{f}$ and the metallicity of the gas for the case with the fiducial case (with $CFE_{ff}=0.2$) and 2 other models with different CFE values ($CFE_{ff}=0.1$ and $0.3$) i.e., $SFE_{f}=C_{Z}~e^{-\log (Z/Z_{\odot})/\tau_{Z}}$ (Eq.~\ref{eq25}).} 
\label{tab4}
\end{table*}

\begin{table*}
\centering
\begin{tabular}{cccc}
\hline
 $M_{clump}$ & $5 \times 10^{4}$ M$_{\odot}$ & $10^{5}$ M$_{\odot}$ & $2 \times 10^{5}$ M$_{\odot}$  \\ 
 \hline
  $C_{Z}$ & $8.73 \times 10^{-2} \pm 4.40 \times 10^{-4}$ & $9.11 \times 10^{-2} \pm 6.81 \times 10^{-4}$ & $9.35\times 10^{-2} \pm 9.32 \times 10^{-4}$  \\
  $\tau_{Z}$ & $8.82 \times 10^{-1} \pm 4.95 \times 10^{-3}$ & $9.15 \times 10^{-1} \pm 7.97 \times 10^{-3}$ & $9.36 \times 10^{-1} \pm 1.10 \times 10^{-2}$\\
\end{tabular}
\caption{Coefficients and exponents of the relation between the final SFE of the cluster $SFE_{f}$ and the metallicity of the gas for $SFE_{f}= C_{Z}~e^{-\log (Z/Z_{\odot})/\tau_{Z}}$ (Eq.~\ref{eq25}), for various clump masses and for the fiducial CFE of $CFE_{ff}=0.2$.} 
\label{tab5}
\end{table*}

\begin{table*}
\centering
\begin{tabular}{ccccccc}
\hline
  &  $Z=1/10~Z_{\odot}$ & $Z=1/6~Z_{\odot}$ & $Z=1/3~Z_{\odot}$ & $Z=1/2~Z_{\odot}$ & $Z=Z_{\odot}$ & $Z=2~Z_{\odot}$  \\ 
 \hline
  $\gamma$ & $1.004$ & $1.011$ & $1.024$ & $1.033$ & $1.047$  & $1.077$\\
  $C_{m}$    & $0.259$ & $0.186$  & $0.114$ & $0.085$ & $0.052$ & $0.027$\\
\end{tabular}
\caption{Coefficients and exponents of the relation between the clump masses and corresponding stellar cluster masses in our models, $M_{cluster}=C_{m}~M_{clump}^{\gamma}$ for the cases with different metallicities.}
\label{tab6}
\end{table*}

\begin{table*}
\centering
\begin{tabular}{cccc}
\hline
 $CFE_{ff}$ & $0.1$ & $0.2$ & $0.3$  \\ 
 \hline
 $C_{Z,rel}$ & $0.731 \pm 6.7 \times 10^{-3}$ & $1.007  \pm 7.5 \times 10^{-3}$ & $1.204 \pm 8 .01 \times 10^{-3}$  \\
 $\tau_{Z,rel}$ & $9.32 \times 10^{-1} \pm 1.01 \times 10^{-2}$ & $9.15 \times 10^{-1} \pm 7.97 \times 10^{-3}$ & $ 9.01 \times 10^{-1} \pm 6.90 \times 10^{-3}$\\
\end{tabular}
\caption{Coefficients and exponents of the relation between the final SFE of the cluster $SFE_{f}$, normalized to the solar metallicity value for the fiducial case with $CFE_{ff}=0.2$, and the metallicity of the gas for the case with the fiducial case (with $CFE_{ff}=0.2$) and 2 other models with different CFE values ($CFE_{ff}=0.1$ and $0.3$) i.e., $SFE_{rel}=SFE_{f}/SFE_{f}(Z=Z_{\odot},CFE_{ff}=0.2)=C_{Z}~e^{-\log (Z/Z_{\odot})/\tau_{Z}}$ (Eq.~\ref{eq26}).} 
\label{tab7}
\end{table*}

{}

\label{lastpage}


\begin{thebibliography}{}

\bibitem[Adams (1996)] {adams96} Adams, F. C., Fatuzzo, M. 1996, ApJ, 464, 256
\bibitem[Arthur (2004)] {arthur04} Arthur, S. J., Kurtz, S. E., Franco, J., Albarr\'{a}n, M. Y. 2004, ApJ, 608, 282
\bibitem[Asplund (2005)] {asplund05} Asplund, M., Grevesse, N., Sauval, A. J. 2005, in ASP Conf. Ser. Vol. 336, Cosmic Abundances as Records of Stellar Evolution and Nucleosynthesis, ed. T. G. Barnes III \& F. N. Bash (San Francisco, CA:ASP), 25
\bibitem[Bastian (2006)] {bastian06} Bastian, N., Goodwin, S. P. 2006, MNRAS, 369, 9 
\bibitem[Basu (2004)] {basu04} Basu, S., Jones, C. E. 2004, MNRAS, 347, L47 
\bibitem[Baumgardt (2003)] {baumgardt03} Baumgardt, H., Kroupa, P. 2007, MNRAS, 380, 1589 
\bibitem[Baumgardt (2008)] {baumgardt08} Baumgardt, H., Kroupa, P., Parmentier, G. 2008, MNRAS, 384, 1231
\bibitem[Bekki (2011)] {bekki11} Bekki, K. 2011, MNRAS, accepted, (arXiv:1011.5956) 
\bibitem[Boily (2003a)] {boily03a} Boily, C. M., Kroupa, P. 2003a, MNRAS, 338, 665
\bibitem[Boily (2003b)] {boily03b} Boily, C. M., Kroupa, P. 2003b, MNRAS, 338, 673  
\bibitem[Boissier (2001)] {boissier01} Boissier, S., Boselli, A. Prantzos, N., Gavazzi, G. 2001, MNRAS, 321, 733 
\bibitem[Braine (2001)] {braine01} Braine, J., Duc, P.-A., Lisenfeld, U., Charmandaris, V., Vallejo, O., Leon, S., Brink, E. 2001, A\&A, 378, 51
\bibitem[Bresolin (2002)] {bresolin02} Bresolin, F., Kennicutt, R. C. 2002, ApJ, 572, 838
\bibitem[Bresolin (2004)] {bresolin04} Bresolin, F., Kudritzki, R. P. 2004, in Origin and Evolution of the Elements, ed. A. McWilliam \& M. Rauch (Cambridge Univ. Press), 283
\bibitem[Canto (2000)] {canto00} Cant\'{o}, J., Raga, A. C., Rodr\'{i}guez, L. F. 2000, ApJ, 536, 896  
\bibitem[Caselli (1995)] {caselli95} Caselli, P., Myers, P. C. 1995, ApJ, 446, 665
\bibitem[Castor (1975)] {castor75} Castor, J., McRay, R., Weaver, R. 1975, ApJ, 200, L107
\bibitem[Chandar (2010)] {chandar10} Chandar, R., Fall, S. M., Whitmore, B. C. 2010, ApJ, 711, 1263 
\bibitem[Chevalier (1985)] {chevalier85} Chevalier, R. A., Clegg, A. W. 1985, Nature, 317, 44
\bibitem[Ciolek (2001)] {ciolek01} Ciolek, G. E., Basu, S. 2001, ApJ, 547, 272 
\bibitem[Clark (2004)] {clark04} Clark, P. C., Bonnell, I. A. 2004, MNRAS, 347, L36 
\bibitem[Cohen (1979)] {cohen79} Cohen, M., Kuhi, L. V. ApJ, ApJS, 41, 743 
\bibitem[Csengeri (2011)] {csengeri11} Csengeri, T., Bontemps, S., Schneider, N., Motte, F., Dib, S. 2011, A\&A, 527, 135
\bibitem[Dib (2006)] {dib06} Dib, S., Bell, E., Burkert, A. 2006, ApJ, 638, 797
\bibitem[Dib (2007e)]{dib07e} Dib, S. 2007, JKAS, 40, 157
\bibitem[Dib (2007f)] {dib07f} Dib, S., Kim, J. 2007, in Small Ionized and Neutral Structures in the Diffuse Interstellar Medium, ed. M. Haverkorn, \& M. Goss (ASP Conference Series), p. 166
\bibitem[Dib (2007a)]{dib07a} Dib, S., Kim, J., V\'{a}zquez-Semadeni, E., Burkert, A., Shadmehri, M. 2007a, ApJ, 661, 262 
\bibitem[Dib (2007b)]{dib07b} Dib, S., Kim, J., Shadmehri, M.  2007b, MNRAS, 381, L40
\bibitem[Dib (2008a)]{dib08a} Dib, S., Brandenburg, A., Kim, J., Maheswar, G., Andr\'{e}, P.  2008a, ApJ, 678, L105 
\bibitem[Dib (2008b)]{dib08b} Dib, S., Shadmehri, M., Maheswar, G., Kim, J., Henning, Th.  2008b, in Massive Star Formation: Observations confront Theory, ASP Conf. Series, Ed. H.  Beuther, H. Linz, T. Henning, 387, 282 
\bibitem[Dib (2008c)] {dib08c} Dib, S., Galv\'{a}n-Madrid, R., Kim, J., V\'{a}zquez-Semadeni, E. 2008c, in the proceedings of the Annual meeting of the French Society of Astrobomy \& Astrophysics, 309, Eds. C. Charbonnel, F. Combes, \& R. Samadi, (arXiv:0808.3305) 
\bibitem[Dib (2009)] {dib09} Dib, S., Walcher, C. J., Heyer, M., Audit, E., Loinard, L. 2009, MNRAS, 398, 1201  
\bibitem[Dib (2010a)] {dib10a} Dib, S., Hennebelle, P., Pineda, J. E., Csengeri, T., Bontemps, S., Audit, E., Goodman, A. A. 2010a, ApJ, 723, 425 
\bibitem[Dib (2010b)] {dib10b} Dib, S., Shadmehri, M., Padoan, P., Maheswar, G., Ojha, D. K., Khajenabi, F. 2010b, MNRAS, 405, 401
\bibitem[Dowell (2008)] {dowell08} Dowell, J. D., Buckalew, B. A., Tan, J. C. 2008, AJ, 135, 823 
\bibitem[Duerr (1982)] {duerr82} Duerr, R., Imhoff, C. L., Lada, C. J. 1982, ApJ, 261, 135
\bibitem[Ellison (2008)] {ellison08} Ellison, S. L., Patton, D. R., Simard, L., McConnachie, A. W. ApJL, 672, L107 
\bibitem[Elmegreen (1997)] {elmegreen97} Elmegreen, B. G., Efremov, Y. N.  1997, ApJ, 480, 235 
\bibitem[Evans (2009)] {evans09} Evans, N., J., Dunham, M. M., J{\o}rgensen, J. K. et al. 2009, ApJS, 181, 321
\bibitem[Fall (2010)] {fall10} Fall, S. M., Krumholz, M. R., Matzner, C. D. 2010, ApJ, 710, L142
\bibitem[Fan (2010)] {fan10} Fan, Z., de Grijs, R., Zhou, X. 2010, ApJ, 725, 200
\bibitem[Fellhauer (2005)] {fellhauer05} Fellhauer, M., Kroupa, P. 2005, ApJ, 630, 879 
\bibitem[Fiedler (1992)] {fiedler92} Fieldler, R. A., Mouschovias, T. Ch. 1992, ApJ, 391, 199
\bibitem[Fukui (1991)] {fukui91} Fukui, Y., Mizuno, A. 1991, in IAU Symp. 147, Fragmentation of Molecular Clouds and Star Formation, eds. E. Falgarone, F. Boulanger, \& G. Duvert, (Dordrecht: Kluwer), 275 
\bibitem[Gail (1979)] {gail79} Gail, H. P., Sedlmayr E. 1979, A\&A, 77, 165 
\bibitem[Galvan (2007)] {galvan07} Galv\'{a}n-Madrid, R., V\'{a}zquez-Semadeni, E., Kim, J., Ballesteros-Paredes, J. 2000, ApJ, 670, 480
\bibitem[Geyer (2001)] {geyer01} Geyer, M. P., Burkert, A. 2001, 323, 988 
\bibitem[Gratier (2010a)] {gratier10a} Gratier, P., Braine, J., Rodriguez-Fernandez, N. J., Israel, F. P., Schuster, K. F., Brouillet, N., Gardan, E. 2010a, A\&A, 512, 68
\bibitem[Gratier (2010b)] {gratier10b} Gratier, P., Braine, J., Rodriguez-Fernandez, N. J. Schuster, K. F., Kramer, C., Xilouris, E. M., Tabatabaei, F. S., Henkel, C., Corbelli, E., Israel, F. et al. 2010b, A\&A, 522, 3
\bibitem[Harper-Clark (2009)] {harperclarck09} Harper-Clark, E., Murray, N. 2009, ApJ, 693, 1696 
\bibitem[Hatchell (2007)] {hatchell07} Hatchell, J., Fueller, G. A., Richer, J. S., Harries, T. J., Ladd, E. F. 2007, A\&A, 468, 1009
\bibitem[Hennebelle (2008)] {hennebelle08} Hennebelle, P., Chabrier, G. 2008, ApJ, 684, 395
\bibitem[Henney (2007)] {henney07} Henney, W. J., in Diffuse Matter fro Star Forming Regions to Active Galaxies, ed. T. W. Hartquist, J. M. Pittard, \& S. A. E. G. Falle (Dordrecht:Springer), 103
\bibitem[Johnstone (2006b)] {johnstone06b} Johnstone, D., Bally, J. 2006, ApJ, 653, 383
\bibitem[Jessop (2000)] {jessop00} Jessop, N. E., Ward-Thompson, D. 2000, MNRAS, 311, 63 
\bibitem[Kennicutt (1989)] {kennicutt89} Kennicutt, R. C. Jr. 1989, ARA\&A, 36, 189
\bibitem[Kennicutt (1998)] {kennicutt98} Kennicutt, R. C. Jr. 1998, ApJ, 498, 541
\bibitem[Kirk (2005)] {kirk05} Kirk, J. M., Ward-Thompson, D., Andr\'{e}, P. 2005, MNRAS, 360, 1506 
\bibitem[Klessen (2000)] {klessen00} Klessen, R. S., Heitsch, F., Mac Low, M.-M. 2000, ApJ, 535, 887
\bibitem[Kobulnicky (1999)] {kobulnicky99} Kobulnicky, H. A., Kennicutt, R. C. Jr., Pizagno, J. L. 1999, ApJ, 515, 544
\bibitem[Koo (1992a)] {koo92a} Koo, B.-C., McKee, C. F. 1992a, ApJ, 388, 93
\bibitem[Koo (1992b)] {koo92b} Koo, B.-C., McKee, C. F. 1992b, ApJ, 388, 103 
\bibitem[Krumholz (2005)] {krumholz05} Krumholz, M. R., McKee, C. F. 2005, ApJ, 630, 250 
\bibitem[Krumholz (2009)] {krumholz09} Krumholz, M. R., Matzner, C. D. 2009, ApJ, 703, 1352
\bibitem[Lamers (1993)] {lamers93} Lamers, H. J. G. L. M., Leitherer, C. 1993, ApJ, 412, 771
\bibitem[Lada (1984)] {lada84} Lada, C. J., Margulis, M., Dearborn, D. 1984, ApJ, 285, 141 
\bibitem[Lada (1991a)] {lada91a} Lada, E. A., Bally, J., Stark, A. A. 1991a, ApJ, 368, 432
\bibitem[Lada (1991b)] {lada91b} Lada, E. A., DePoy, D. L., Evans, J. H. Gatley, I. 1991b, ApJ, 371, 171 
\bibitem[Lada (2003)] {lada03} Lada, C. J., Lada, E.  A. 2003, ARA\&A, 41, 57
\bibitem[Lada (2010)] {lada10} Lada, C. J., Lombardi, M., Alves, J. F. 2010, ApJ, 724, 687 
\bibitem[Lara-Lopez (2010)] {laralopez10} Lara-L\'{o}pez, M. A., Cepa, J., Bongiovanni, A., P\'{e}rez Garc\'{i}a, A. M., Ederoclite, A., Casta\~{n}eda, H., Fern\'{a}ndez Lorenzo, M., Povi\'{c}, M., Sanchez-Portal, M. 2010, A\&A, 521, L53 
\bibitem[Larsen (2009)] {larsen09} Larsen, S. S. 2009, A\&A, 494, 539 
\bibitem[Lee (1999)] {lee99} Lee, C. W., Myers, P. C. 1999, ApJS, 123, 233
\bibitem[Lee (2005)] {lee05} Lee, J.-K., Rolleston, W. R. J., Dufton, P. L., Ryans, R. S. I . 2005, A\&A, 429, 1025 
\bibitem[Leitherer (1992)] {leitherer92} Leitherer, C., Robert, C., Drissen, L. 1992, ApJ, 401, 596 
\bibitem[Lepine (2008)] {lepine08} L\'{e}pine, S., Moffat, A. F. J. 2008, ApJ, 136, 548
\bibitem[Leroy (2006)] {leroy06} Leroy, A. K., Bolatto, A., Walter, F., Blitz, L. 2006, ApJ, 643, L825 
\bibitem[Leroy (2008)] {leroy08} Leroy, A. K., Walter, F., Brinks, E., Bigiel, F., de Blok, W. J. G., Madore, B., Thornley, M. D. 2008, AJ, 136, 2728 
\bibitem[Li (2006)] {li06} Li, Z.-Y., Nakamura, F. 2006, ApJ, 640, L187
\bibitem[Li (2010)] {li10} Li, Z.-Y., Wang, P., Abel, T., Nakamura, F. 2010, ApJ, 720, L26  
\bibitem[Mannucci (2010)] {manucci10} Mannucci, F., Cresci, G., Maiolino, R., Marconi, A., Gnerucci, A. 2010, MNRAS, 408, 2115
\bibitem[Mandushev (1991)] {mandushev91} Mandushev, G., Staneva, A., Spasova, N. 1991, A\&A, 252, 94
\bibitem[Mathews (1969)] {mathews} Mathews, W. G. 1969, ApJ, 157, 583 
\bibitem[Matzner (2000)] {matzner00} Matzner, C. D., McKee, C. F. 2000, ApJ, 545, 364 
\bibitem[Matzner (2002)] {matzner02} Matzner, C. D. 2002, ApJ, 566, 302 
\bibitem[Matzner (2007)] {matzner07} Matzner, C. D. 2007, ApJ, 659, 1394 
\bibitem[McKee (1984)] {mckee84} McKee, C. F., Van Buren, D., Lazareff, B. 1984, ApJ, 278, L115 
\bibitem[McKee (1987)] {mckee97} McKee, C. F., Williams, J. P. 1997, ApJ, 476, 144 
\bibitem[McKee (1989)] {mckee89} McKee, C. F. 1989, ApJ, 345, 782
\bibitem[Meynet (2000)] {meynet00} Meynet, G., Maeder, A. 2000, A\&A, 361, 101
\bibitem[Morel (1997)] {morel97} Morel, P. 1997, A\&AS, 124, 597 
\bibitem[Morel (2008)] {morel08} Morel, P., Lebreton, Y. 2008, Ap\&SS, 316, 61 
\bibitem[Mueller (2002)] {mueller02} Mueller, K. E., Shirley, Y. L., Evans, N. J. II, Jacobson, H. R. 2002, ApJS, 143, 469 
\bibitem[Murgia (2002)] {murgia02} Murgia, M., Craspi, A., Moscadelli, L., Gregorini, L. 2002, A\&A, 385, 412
\bibitem[Murray (2010)] {murray10} Murray, N., Quataret, E., Thompson, T. A. 2010, ApJ, 709, 191 
\bibitem[Myers (1986)] {myers86} Myers, P. C., Dame, T. M., Thaddeus, P., Cohen, R. S., Silverberg, R. F. 1986, ApJ, 301, 398 
\bibitem[Nakamura (2005)] {nakamura05} Nakamura, F., Li, Z.-Y. 2005, ApJ, 631, 411 
\bibitem[Nakamura (2007)] {nakamura07} Nakamura, F., Li, Z.-Y. 2007, ApJ, 662, 395
\bibitem[Olmi (2002)] {olmi02} Olmi, L., Tesi, L. 2002, A\&A, 392, 1053
\bibitem[Padoan (1995)] {padoan95} Padoan, P., 1995, MNRAS, 277, 377 
\bibitem[Padoan (2002)] {padoan02} Padoan, P., Nordlund, \AA. 2002, ApJ, 576, 870
\bibitem[Padoan (2011)] {padoan11} Padoan, P., Nordlund, \AA. 2011, ApJ, 730, 40  
\bibitem[Pandey (1990)] {pandey90} Pandey, A. K., Paliwal, D. C., Mahra, H. S. 1990, ApJ, 362, 165 
\bibitem[Parmentier (2007)] {parmentier07} Parmentier, G., Gilmore, G. 2007, MNRAS, 377, 352 
\bibitem[Parmentier (2008)] {parmentier08} Parmentier, G., Goodwin, S. P., Kroupa, P., Baumgardt, H. 2008, ApJ, 678, 347
\bibitem[Parmentier (2009)] {parmentier09} Parmentier, G., Fritze, U. 2009, ApJ, 690, 1112
\bibitem[Parmentier (2011)] {parmentier11} Parmentier, G. 2011, MNRAS, 218, (arXiv:1101.0813)
\bibitem[Pauldrach (1990)] {pauldrach90} Pauldrach, A. W. A., Kudritzki, R. P., Puls, J., Butler, K. 1990, A\&A, 228, 125 
\bibitem[Perez-Montero (2005)] {perezmontero05} P\'{e}rez-Montero, E., D\'{i}az, A. I.  2005, MNRAS, 361, 1063
\bibitem[Piau (2011)] {piau11} Piau, L., Kervella, P., Dib, S., Hauschildt, P. 2011, A\&A, 526, 100 
\bibitem[Price (2008)] {price08} Price, D. J., Bate, M. R. 2008, MNRAS, 385, 1820
\bibitem[Prinja (2010)] {prinja10} Prinja, R. K., Massa, D. L. 2010, A\&A, 521, L55 
\bibitem[Rengarajan (1984)] {rengarajan84} Rengarajan, T. N. 1984, ApJ, 287, 671
\bibitem[Rolleston (2002)] {rolleston02} Rolleston, W. R. J., Trundle, C., Dufton, P. L. 2002, A\&A, 396, 53 
\bibitem[Rolleston (2003)] {rolleston03} Rolleston, W. R. J., Venn, K., Tolstoy, E., Dufton, P. L. 2003, A\&A, 400, 21  
\bibitem[Rosas (2010)] {rosas10} Rosas-Guevara, Y., V\'{a}zquez-Semadeni, E., G\'{o}mez, G. C., Jappsen, A.-K. 2010, MNRAS, 406, 1875   
\bibitem[Rownd (1999)] {rownd99} Rownd, B. K., Young, J. S. 1999, AJ, 118, 670 
\bibitem[Russell (1992)] {russell92} Russell, S. C., Dopita, M. A. 1992, ApJ, 384, 508
\bibitem[Saito (2007)] {saito07} Saito, H., Saito, M., Sunada, K., Yonekura, Y. 2007, ApJ, 659, 459
\bibitem[Scheepmaker (2007)] {scheepmaker07} Scheepmaker, R. A., Haas, M. R., Gieles, M., Bastian, N., Larsen, S. S., Lamers, H. J. G. L. M. 2007, A\&A, 469, 925
\bibitem[Schiminovitch (2010)] {schiminovitch10} Schiminovitch, D., Catinella, B., Kauffman, G., Fabello, S., Wang, J. et al. 2010, MNRAS, 408, 919 
\bibitem[Shadmehri (2004)] {shadmehri04} Shadmehri, M. 2004, MNRAS, 354, 373 
\bibitem[Schaerer (2011)] {schaerer11} Schaerer, D., Charbonnel, C. 2011, MNRAS, accepted, (arXiv:1101.1073) 
\bibitem[Shull (1980)] {shull80} Shull, J. M. 1980, ApJ, 238, 860 
\bibitem[Smith (2002)] {smith02} Smith, V., Hinkle, K. H., Cunha, K., Plez, B., Lambert, D. et al. 2002, AJ, 124, 3241
\bibitem[Stevens (2003)] {stevens03} Stevens, I. R., Hartwell, J. M. 2003, MNRAS, 339, 280 
\bibitem[Vazquez-Semadeni (1999)] {vazquez99} V\'{a}zquez-Semadeni, E., Passot, T. 1999, In interstellar Turbulence, eds.,  J. Franco and A Carrami\~{n}ana, University Press (Cambridge), p. 223
\bibitem[Vazquez (2003)] {vazquez03} V\'{a}zquez-Semadeni, E., Ballesteros-Paredes, J., Klessen, R. S. 2003, ApJ, 585, L131
\bibitem[Vazquez (2005a)] {vazquez05a} V\'{a}zquez-Semadeni, E., Kim, J., Shadmehri, M., Ballesteros-Paredes, J. 2005a, ApJ, 618, 344
\bibitem[Vazquez (2005b)] {vazquez05b} V\'{a}zquez-Semadeni, E., Kim, J., Ballesteros-Paredes, J. 2005b, ApJ, 630, L49
\bibitem[Vazquez (2010)] {vazquez10} V\'{a}zquez-Semadeni, E., Col\'{i}n, P., G\'{o}mez, G. C., Ballesteros-Paredes, J., Watson, A. 2010, ApJ, 715, 1302  
\bibitem[Veltchev (2011)] {veltchev11} Veltchev, T. V., Klessen, R. S., Clark, P. C. 2011, MNRAS, 411, 301 
\bibitem[Vink (1999)] {vink00} Vink, J. S., de Koter, A., Lamers, H. J. G. L. M. 1999, A\&A, 350, 181 
\bibitem[Vink (2000)] {vink00} Vink, J. S., de Koter, A., Lamers, H. J. G. L. M. 2000, A\&A, 362, 295
\bibitem[Vink (2001)] {vink01} Vink, J. S., de Koter, A., Lamers, H. J. G. L. M. 2001, A\&A, 369, 574 
\bibitem[Ward-Thompson (2007)] {wardthompson07} Ward-Thompson, D., Andr\'{e}, P., Crutcher, R., Johnstone, D., Onishi, T., Wilson, C. 2007, in Protostars and Planets V., ed.  B. Reipurth, D. Jewitt, \& K., Keil (Tuscon:Univ. of Arizona Press), 33
\bibitem[Warin (1996)] {warin96} Warin, S., Castets, A., Langer, W. D., Wislon, R. W., Pagani, 1996, A\&A 306, 935
\bibitem[Weaver (1977)] {weaver77} Weaver, R., McRay, R., Castor, J., Sharpir, P., Moore, R. 1977, ApJ, 218, 377
\bibitem[Whitworth (2001)] {whitworth01} Whitworth, A. P., Ward-Thompson, D. 2001, ApJ, 547, 317
\bibitem[Wilking (1983)] {wilking83} Wilking, B. A., Lada, C. J. 1983, ApJ, 274, 698 
\bibitem[Wolf (1990)] {wolf90} Wolf, G., Lada, C. J., Bally, J. 1990, AJ, 100, 1892
\bibitem[Young (1986)] {young86} Young, J. S., Schloerb, F. P. Kenney, J. D., Lord, S. D. 1986, ApJ, 304, 443 
\bibitem[Young (1996)] {young96} Young, J. S., Allen, L., Kenney, J. D. P., Lesser, A., Rownd, B. 1996, AJ, 112, 1903  
\bibitem[Zaritsky (1994)] {zaritsky94} Zaritsky, D., Kennicutt, R. C. Jr., Huchra, J. P. 1994, ApJ, 420, 87
\bibitem[Zasov (2006)] {zasov06} Zasov, A. V., Abramova, O. V. 2006, Astr. Rep. 50., 874
\bibitem[Zhang (1999)] {zhang99} Zhang, Q., Fall, S. M. 1999, ApJ, 527, L81 
 
\end{thebibliography}
\end{document}